\documentclass[12pt]{article}
   \usepackage{epsfig}
      \usepackage{amsmath}
      \usepackage[normalem]{ulem}
      \usepackage{amssymb}
      \usepackage{caption}
      \usepackage{tensor}
      \usepackage{braket}
      \usepackage{mathtools}
\usepackage{subcaption}
\usepackage{cite}

\DeclarePairedDelimiter\floor{\lfloor}{\rfloor}

\usepackage{epstopdf}

\newcommand{\calh}{\mathcal{H}}
\newcommand{\calI}{\mathcal{I}}
\newcommand{\calJ}{\mathcal{J}}
\newcommand{\calu}{\mathcal{U}}

\usepackage{amsmath}

\usepackage{setspace}
  \usepackage{graphicx}
  \usepackage{color}

 \newcommand{\be}{\begin{equation}}
\newcommand{\bea}{\begin{eqnarray}}
\newcommand{\eea}{\end{eqnarray}}
\newcommand{\beq}{\begin{equation}}
 \newcommand{\ee}{\end{equation}}

\def\sltr{$\widetilde{SL}(2,\mathbb{R})$ }

\begin{document}
  \renewcommand{\theequation}{\thesection.\arabic{equation}}

\begin{titlepage}

  \bigskip\bigskip\bigskip\bigskip\bigskip

  \bigskip

  \vspace*{100px}

\centerline{\Large \bf {Entanglement Entropy in Jackiw-Teitelboim Gravity}}

    \bigskip

  \begin{center}

 \bf { Daniel Louis Jafferis and David K. Kolchmeyer}
  \bigskip \rm
\bigskip

{\it  Center for the Fundamental Laws of Nature, Harvard University, Cambridge, MA, USA}
\smallskip

\vspace{1cm}
  \end{center}

  \bigskip\bigskip

 \bigskip\bigskip
  \begin{abstract}

We compute the entanglement entropy and Renyi entropies of arbitrary pure states in pure Jackiw-Teitelboim gravity in Lorentz signature. We apply the quantum Hubeny-Rangamani-Ryu-Takayanagi  formula by computing the quantum corrected area term and the bulk entropy term. The sum of these two terms for the Hartle-Hawking state agrees with the black hole entropy above extremality computed from the Euclidean disk path integral. We interpret the area term as the universal contribution of a defect operator that plays a crucial role in our Lorentzian interpretation of the Euclidean replica trick in gravity.

 \medskip
  \noindent
  \end{abstract}

  \end{titlepage}

  \tableofcontents

\section{Introduction}

The most striking aspect of entropy in gravitational theories
is that it is UV finite and captured by the long distance effective theory of general relativity. In particular, the famous formula $S \approx {\cal A}/4G_N$ relates a microscopic entanglement entropy to the area of the extremal surface that divides space into two regions. This should be contrasted with the conventional situation in which long distance effective theories lack the appropriate data.

Entanglement entropy is defined with respect to a factorization of the system, and in the context of the AdS/CFT duality a precise notion is provided by considering the dual of subregions on the boundary. In the semiclassical limit of the bulk, the dual of a boundary subregion is a bulk subregion bounded by the codimension-2 Hubeny-Rangamani-Takayanagi (HRT) surface \cite{HRT}, namely the bulk extremal surface of minimal area that is homologous to the boundary subregion. Note that there is no obvious way to assign an entanglement entropy to a general bulk region because there will not be an intrinsic diffeomorphism invariant way of specifying the location of the bounding surface.

In this paper, we will give a new perspective on the Lorentzian interpretation of entanglement entropy in gravity, by carefully considering how a factorized Hilbert space can be defined and relating it to path integral calculations of Renyi entropies. We will do this in the toy model of Jackiw-Teitelboim gravity in 1+1 dimensions. The situations we consider will have boundary subregions that consist of an entire component of a disconnected boundary, as arises in two sided black hole spacetimes. 

The discovery of the Hubeny-Rangamani-Ryu-Takayanagi formula \cite{RT,HRT} led to computations of entanglement entropy in quantum gravity that have greatly furthered our understanding of how bulk physics emerges from the boundary in AdS/CFT.
Perturbative quantum corrections to the HRT formula have been addressed in \cite{FLM,EW,alliteration}. The conjecture is that the quantum-corrected entropy should be computed with respect to the quantum extremal surface instead of the classical extremal surface. A quantum extremal surface is a codimension-2 surface that extremizes the generalized entropy $S_{\text{gen}}$, which is defined by
\begin{equation} S_{\text{gen}} \equiv \frac{\braket{A}}{4 G_N} + S_{\text{bulk}}, \label{eq:sgen} \end{equation}
where the first term is the expectation value of the area operator, and the second term is the entanglement entropy of bulk fields across the surface, including gravitons. While this formula has been used in various contexts \cite{QuantumPenrose,SurfaceTheory,pagecurve1,pagecurve2}, it is not fully understood how to define the entanglement entropy of gravitons beyond leading order \cite{JLMS,quantumfocussing}. 

Quantum corrections to gravitational entanglement entropy are crucial for understanding bulk reconstruction \cite{HarlowRT,alphabits,universalrecovery}, yet concrete calculations have mostly ignored the effects of graviton entanglement by working in the semiclassical limit \cite{pagecurve1,pagecurve2}. Interesting studies of graviton entanglement entropy include \cite{freegravitons}, which considers free gravitons, and \cite{livingontheedge}, which discretizes spacetime. Our understanding of entanglement entropy in quantum gravity is incomplete.

The perturbatively corrected HRT formula can be proved \cite{LM,FLM,alliteration} using the replica trick by evaluating the Renyi entropies with the Euclidean gravity path integral. The origin of the UV finiteness and universality of entropy in gravity can easily be seen from that perspective. Moreover, there are no ambiguities in the definition of the contribution of gravitational fluctuations to all orders in perturbation theory. The main goal of our work is to relate such calculations to a Lorentzian description of the entanglement. 

To study gravitational entanglement entropy in a controllable setting, we turn to Jackiw-Teitelboim (JT) gravity \cite{Jackiw,Teitelboim,MSY}. JT gravity is a two-dimensional effective theory of quantum gravity that is part of the IR limit of the SYK model \cite{FactorizationProblem}, allowing for more explicit studies of the bulk-boundary correspondence. In this paper, we will consider pure JT gravity, without matter. Unlike effective theories of gravity in four or more dimensions, JT gravity is a sensible, self-consistent quantum theory even if we choose to not impose a UV cutoff scale.\footnote{Because JT gravity is self-consistent at all energy scales, we must be especially careful not to apply it outside of its domain of validity. We will interpret JT gravity as an effective theory, not as a full theory of quantum gravity.} Despite its relative simplicity, JT gravity has proven to be a useful toy model for developing our understanding of AdS/CFT \cite{mertensbhevaporation, MSY,zhenbinyang,clocksandrods,holographiccomplexity, pagecurve1, eternaltraversable, symmetriesnearhorizon, sss, semiclassicalramp, saad}.

For any pure state, we compute the entropy of entanglement between the two asymptotically nearly-AdS$_2$ boundaries. We are particularly interested in computing the entanglement entropy of the Hartle-Hawking state, which describes the nearly AdS$_2$ eternal black hole prepared by the Euclidean gravity path integral on the half disk.	 The entanglement entropy of the Hartle-Hawking state is already known from the computation of the Euclidean path integral on the disk \cite{stanfordwitten}. For inverse temperature $\beta$, the Euclidean calculation tells us that the  entropy (above extremality) is given by
\begin{equation}
\label{eq:introhartlehawkingentropy}
S_{HH}(\beta) = \frac{\pi \phi_b}{2  G_N \beta}- \frac{3}{2}\log \frac{16 \pi G_N \beta}{\phi_b} + \frac{3}{2}+\log(4 \pi^{3/2}) .
\end{equation}
In the semiclassical limit $G_N \rightarrow 0$, the leading contribution is
\begin{equation}
\label{eq:semiclassicalentropy}
\frac{\pi \phi_b}{2  G_N \beta},
\end{equation}
which is the horizon value of the dilaton divided by $4 G_N$. The origin of this semiclassical area term has been studied in \cite{JenniferLin,JTdefects}. 

We will present two main results. Firstly, we describe a particular factorization map that reproduces the $S_{\text{bulk}}$ contribution to the right hand side of \eqref{eq:sgen} in the Lorentzian theory for the Hartle-Hawking state. The factorization map appears naturally in relating JT gravity to the Schwarzian description of the boundaries. In Lorentzian JT gravity, the expectation value of the area operator can be defined without factorizing the Hilbert space. Upon summing the area term and the $S_{\text{bulk}}$ term for the Hartle-Hawking state, we find agreement with \eqref{eq:introhartlehawkingentropy}.\footnote{The entropy we compute also contains a universal, infinite additive constant. However, a counterterm in the Euclidean path integral can shift \eqref{eq:introhartlehawkingentropy} by an additive constant. We are not interested in any additive constants that can be absorbed into the definition of the extremal entropy.} Thus, we show how to apply the quantum HRT formula to pure JT gravity.

Secondly, we give a Lorentzian interpretation of the Euclidean path integral calculation of the Renyi entropies. We will see that topological obstructions prevent us from using a local boundary condition to define an isometric cutting map for the gravity Hilbert space. Thus, we need to modify the usual trace formula for the Renyi partition functions ($Z[n] = \text{Tr } \rho^n$). For the Hartle-Hawking state, the modified trace formula is \begin{equation} \label{eq:modifiedtrace} Z[n] = \text{Tr } D e^{- n \beta H },  \end{equation} where $D$ is an operator that does not depend on $\beta$.\footnote{This formula corresponds to \eqref{eq:partitionfunctionwithdefectoperator} in Section \ref{sec:splitting}.} We show in Section \ref{sec:splitting} that $D$ is Hermitian and that it commutes with the Hamiltonian. This modification will lead to a more precise characterization of the area term, as the expectation value of a universal operator acting in a factorized Hilbert space of the effective theory. The entanglement entropy of degrees of freedom in the factorized effective theory is $S_{\text{bulk}}$, while the extra insertion of the operator $D$ results from degrees of freedom in the full UV theory whose universal effect in the long distance theory is to implement the correct topological restrictions.

In the next section, we explain the overall logic of this paper. We briefly summarize the next section as follows. To compute entanglement entropy in JT gravity, we first define a factorized JT gravity Hilbert space, $\calh_L \otimes \calh_R$, by using a local boundary condition placed on a codimension-1 ``brick wall'' boundary. The boundary condition we use is easy to state in the BF formulation of the theory. As we will explain, it is not possible to use a local boundary condition to define an \emph{isometric} cutting map (as in Figure \ref{fig:cuttingfigure}). However, we argue that there is only one physically reasonable choice of factorization map from the unfactorized JT gravity Hilbert space $\calh$ into $\calh_L \otimes \calh_R$, which we call $\calJ : \calh \rightarrow \calh_L \otimes \calh_R$. By studying the Euclidean path integral, we argue that $\calJ$ must take the form $\calJ = \sqrt{D} \calI$, where $\calI : \calh \rightarrow \calh_L \otimes \calh_R$ is a cutting map, and $D$ is a simple Hermitian operator that acts on the factorized Hilbert space. The map $\calI$ is not isometric, while $\calJ$ is:
\begin{equation}
\calJ^\dagger \calJ = \calI^\dagger D \calI = 1.
\end{equation}
In the BF formulation of the Euclidean JT theory, a smooth metric corresponds to a flat gauge connection with a defect \cite{JTdefects}. The operator $D$ is the defect operator in the Lorentzian theory. This explains why we insert $D$ only once into the modified trace formula \eqref{eq:modifiedtrace} for the $n$th Renyi partition function. Using the modified trace formula, we find that the defect operator makes a contribution to the entanglement entropy that precisely matches the quantum area term in the HRT formula.

The remainder of this paper is summarized as follows. In Section \ref{sec:JTgravity}, we review JT gravity and establish our conventions. In particular, we show how JT gravity may be naturally described as two copies of the Lorentzian Schwarzian theory subject to certain constraints. In Section \ref{sec:schwarziantheory}, we quantize the Schwarzian theory in Lorentz signature. In Section \ref{sec:splitting}, we describe how we isometrically map the Hilbert space of JT gravity to the tensor product Hilbert space of two Schwarzian theories (that is, we define $\calJ$). In Section \ref{sec:defect}, we show how the Euclidean Schwarzian theory (described by a particle propagating near the AdS boundary) follows from imposing a local boundary condition on a brick wall in the Euclidean gravity path integral. In Section \ref{sec:euclidpathhh}, we show how the Euclidean Schwarzian path integral can be used to compute the image of the Hartle-Hawking state under the factorization map described in Section \ref{sec:splitting}. In Section \ref{sec:ee}, we compute the entanglement entropy of arbitrary states by summing the quantum-corrected area term and the bulk entropy term in the HRT formula. We also compute the Renyi partition functions from the modified trace formula with the defect operator. We explicitly check that the defect operator makes a universal contribution to the entanglement entropy that agrees with the expectation value of the area operator. 

\section{Main Results}

\label{sec:mainresults}

The purpose of this section is to fully explain the logic and motivations underlying our calculations. In particular, we comment on how we factorize the Hilbert space and how we interpret the Euclidean path integral. 

\subsection{Factorizing the Hilbert space of JT  gravity}

In this subsection, we elaborate on the process of factorizing a Hilbert space in either quantum field theory or quantum gravity. Then, we describe our procedure for factorizing the JT gravity Hilbert space. We emphasize that there is a difference between using a local boundary condition to define a factorized Hilbert space and using a local boundary condition to define a map from the original Hilbert space to the factorized Hilbert space. In this paper, we use a local boundary condition to define a factorized Hilbert space for JT gravity, but the factorization map we use cannot be produced by a local boundary condition in the path integral.

The entanglement entropy of a state is defined as the von Neumann entropy of the reduced density matrix $\rho$:
\begin{equation}
S = - \text{Tr } \rho \log \rho.
\label{eq:sbulkvonneumann}
\end{equation}
In particular, we use \eqref{eq:sbulkvonneumann} to compute $S_{\text{bulk}}$ in \eqref{eq:sgen}. The main obstacle to determining $\rho$ is the fact that the Hilbert space of the theory in question is usually not factorized, as was shown in \cite{FactorizationProblem} for JT gravity. We must map the Hilbert space into a factorized Hilbert space with a linear and isometric factorization map. The fundamental problem is that different choices of the factorization map will produce different results for \eqref{eq:sbulkvonneumann}. For instance, we can map an energy eigenstate $\ket{E}$ into an entangled state in a factorized Hilbert space as follows:\footnote{For simplicity, we have written this equation pretending that $\ket{E}$ is normalized to unity. In JT gravity, the energy eigenstates $\ket{E}$ are only delta-function normalizable. Also, the image of the factorization map given here is a proper subspace of a factorized Hilbert space.}
\begin{equation}
\label{eq:generalmap}
\calu: \ket{E} \rightarrow \sum_{m = 1}^{d(E)} \frac{1}{\sqrt{d(E)}}\ket{E \, m} \otimes \ket{E \, m},
\end{equation}
where we are free to specify the degeneracy function $d(E)$. Since one can achieve virtually any result for \eqref{eq:sbulkvonneumann} with an appropriate choice of factorization map, it is important to identify the general principles that select one choice of factorization map over another. Otherwise, we cannot compute entanglement entropy in a meaningful way.

A sensible requirement for the factorization map $\calu$  is that it preserves locality. A relatively weak notion of locality can be formulated algebraically. Namely, that the algebra of local operators on the left (respectively right) of the entangling surface should act only on the left (resp. right) tensor factor under the action of the cutting map. However, in pure JT gravity, this is always obeyed by \eqref{eq:generalmap} because the left and right algebras in pure JT gravity are simply generated by the left and right Hamiltonia (which moreover are equal by the Gauss law). Other operators in JT gravity, such as the length of the spatial geodesic at given boundary times, involve both boundaries and thus are in neither the left nor right algebra. 

\begin{figure}
	\centering
	\includegraphics[width=0.7\linewidth]{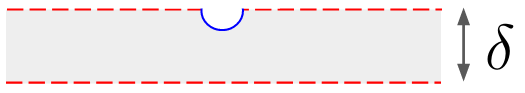}
	\caption{We illustrate how to use a local boundary condition to define a cutting map from an unfactorized Hilbert space to a factorized Hilbert space. Given a state in the unfactorized Hilbert space, we can evolve for a small time $\delta$ with the Euclidean path integral with an additional boundary inserted, shown in blue. The boundary condition at the blue boundary defines each of the Hilbert space factors. It also defines the cutting map. The boundary conditions on the dashed red boundaries correspond to defining states in the Hilbert space. In quantum field theory, we additionally need to specify the size and shape of the blue boundary. In quantum gravity, the geometry is dynamical, so it is enough to only specify the boundary condition. The time parameter $\delta$ should be taken to zero. We leave $\delta$ implicit in the remainder of this paper. The blue boundary is often referred to as a brick wall.}
	\label{fig:cuttingfigure}
\end{figure}

A much stronger notion of locality would be implemented by requiring that the map $\calu$ is obtained from the path integral by splitting the spacetime with a local boundary condition \cite{Tachikawa}. We call such a factorization map a ``cutting map.'' We illustrate the process of using a boundary condition to define a cutting map in Figure \ref{fig:cuttingfigure}. In quantum field theory, such a boundary condition could include local background terms involving the intrinsic metric and the extrinsic curvature \cite{DJ}. However, in gravity, the metric is dynamical, so there are no such non-dynamical background terms. 

The aforementioned algebraic locality condition is insufficient even in quantum field theory \cite{DJ}. For example, in pure Chern-Simons theory on $S^2$, the algebra of local operators is empty, and one can obtain any result for the entanglement between the two hemispheres by splitting the unique state analogously to one term in \eqref{eq:generalmap}. However, requiring that the factorization map results from a local boundary condition in the path integral leads uniquely to the usual non-trivial result \cite{GabrielWong}.
It will turn out that this path integral procedure will have to be modified in gravity, in a way that will give a new interpretation of the quantum HRT formula.

Before we explain the factorization map that we use in pure JT gravity, we first briefly comment on pure 2D BF theory. In BF theory, one may define a cutting map by requiring that the boundary component of the gauge field vanishes at the brick wall boundary. The states in the factorized Hilbert space  need not  satisfy the Gauss law constraint at the entangling surface. We will sometimes refer to the the image of the original Hilbert space in the factorized Hilbert space as the physical or gauge invariant states. Those do obey the Gauss law constraint. Of course, in the factorized system with a brick wall boundary, there is nothing unphysical or non-gauge invariant about the rest of the states. While the presence of extra states may seem unsettling, we emphasize that the structure of the factorized Hilbert space naturally arises from a simple local boundary condition on the gauge field. 

To factorize the Hilbert space of JT gravity, we can apply our knowledge of factorizing the Hilbert space of BF theory. In Section \ref{sec:defect}, we rewrite the action of pure Euclidean JT gravity in BF variables (which is equivalent to the first-order formalism) and consider the path integral of the theory quantized on a spatial interval times Euclidean time. On one boundary, we place the usual asymptotically nearly-AdS$_2$ boundary condition associated with JT theory. On the other boundary, we impose that the boundary component of the frame fields and spin connection vanishes:
\begin{equation}
\left. e^a \right|_{\text{boundary}} = \left. \omega \right|_{\text{boundary}} = 0, \quad \quad a \in \{1,2\}.
\label{eq:introbc}
\end{equation}
Given these boundary conditions, we show in Section \ref{sec:defect} that the path integral is formally equivalent to that of the Euclidean Schwarzian theory,\footnote{Whenever we use the term ``Schwarzian theory'' we are referring to the Lorentzian Schwarzian theory of Section \ref{sec:schwarziantheory}, which is not to be confused with the Euclidean Schwarzian theory of Section \ref{sec:euclidpathhh}.} whose dynamical degrees of freedom are described by a particle propagating near the boundary of the Euclidean hyperbolic disk \cite{kitaevsuh,zhenbinyang}. We find the boundary particle formalism of JT gravity to be convenient for our purposes, especially because it naturally follows from placing \eqref{eq:introbc} on an inner boundary. 

We are free to impose \eqref{eq:introbc} on the brick wall boundary to define a cutting map. For now, we will only be interested in the structure of the factorized Hilbert space and not the map itself. The previous paragraph implies that each factor of the factorized Hilbert space is given by the Hilbert space of the Euclidean Schwarzian theory. There is a subtlety, however. We have to take into account the fact that the Euclidean and Lorentzian gravity theories are defined with different contours for the fields. In particular, as we show in Section \ref{sec:euclidpathhh}, the Hamiltonian of the Euclidean Schwarzian theory is not Hermitian, while the Hamiltonian of the Lorentzian Schwarzian theory is Hermitian. We are ultimately interested in Lorentzian JT gravity, but we require the Euclidean path integral to define a cutting map as in Figure \ref{fig:cuttingfigure}. Because we choose to formulate JT gravity as a theory of boundary particles (which is a consequence of our desire to specify the brick wall boundary condition in the first-order formalism), we find that wavefunctions prepared using the Euclidean Schwarzian path integral must be analytically continued in an appropriate way before they can be interpreted as wavefunctions of the Lorentzian theory. This is due to the fact that the target space of the boundary particle is the hyperbolic disk in the Euclidean Schwarzian theory and global AdS$_2$ in the Lorentzian Schwarzian theory. We work out the appropriate analytic continuation explicitly for the Hartle-Hawking wavefunction in JT gravity in Section \ref{sec:euclidpathhh}. The upshot is that the factorized Hilbert space of the Lorentzian JT theory that we use in this paper is given by two copies of the Lorentzian Schwarzian theory, each of which describes a particle propagating near a boundary of global AdS$_2$. While this method of defining a factorized Hilbert space arises from a local boundary condition as desired, we note that the factorization map we ultimately use is \emph{not} the cutting map defined by \eqref{eq:introbc}. However, it is related to the cutting map in a simple way.

In fact, there is only one physically sensible choice of the factorization map from the unfactorized Hilbert space of Lorentzian JT gravity to the factorized Hilbert space described just above. We elaborate on this factorization map in Section \ref{sec:splitting}. For now, we mention that just as in BF theory, the physical states in the factorized JT gravity Hilbert space must satisfy certain constraints, which we describe in Section \ref{sec:JTgravity}. The linear factorization map that we use simply maps an energy eigenstate of JT gravity with energy $E$ to the unique energy eigenstate of the factorized Hilbert space that obeys the constraints and has energy $E$. Our factorization map is manifestly an isometry.

After factorizing the Hilbert space of JT gravity, the thermofield double state becomes well-defined. Interestingly, our factorization map does \emph{not} map the Hartle-Hawking state to the thermofield double state of two copies of the Schwarzian theory. The reduced density matrix of the Hartle-Hawking state is not proportional to $e^{-\beta H}$. Rather, it is given by\footnote{As shown in Section \ref{sec:schwarziantheory}, the Schwarzian theory contains states of arbitrarily negative energy. In this paper, we restrict ourselves to the positive energy states for reasons given in Section \ref{sec:splitting}. Thus, the trace of $\rho_\beta$ is well-defined. We could define $D$ to annihilate the orthogonal complement of the positive energy subspace of the Hilbert space of the Lorentzian Schwarzian theory.}
\begin{equation}
\label{eq:densitymatrixintro}
\rho_\beta = \frac{D e^{-\beta H}}{\text{Tr } D e^{-\beta H}},
\end{equation}
where $D$ is a defect operator that does not depend on $\beta$.\footnote{The operator $D$ is an operator in the Lorentzian Schwarzian theory. Note that in Section \ref{sec:euclidpathhh} we will define an operator $D_E$ in the Euclidean Scwharzian theory, which also admits a Hilbert space description. While $D$ and $D_E$ have similar interpretations, they are mathematically distinct. We define $D$ in \eqref{eq:defectoperatordefinition} and we define $D_E$ in \eqref{eq:defectoperatordefinitioneuclidean}.} This fact reflects a fundamental tension between defining a factorization map that is both local (in the strong sense described earlier) and isometric. As explained above, we use a local boundary condition to define the factorized Hilbert space, but we do not use a local boundary condition to define the factorization map itself. Rather, we choose our factorization map to be an isometry onto the physical subspace of the factorized Hilbert space. In BF theory, one can define both a factorized Hilbert space and an isometric factorization map with a local boundary condition. In the next subsection, we explain why this is not the case in gravity by examining the Euclidean gravitational path integral. We conclude that in Lorentzian JT gravity, we must introduce an insertion of $D$ into the computation of the disk partition function. This leads to our prescription for computing entanglement entropies in JT gravity for all pure states. We expect that it is possible to generalize our approach to other theories of gravity.

\subsection{A Lorentzian interpretation of the Euclidean gravity path integral}

In non-gravitational theories, the Renyi partition function $Z[n]$ is computed on the $n$-fold cover of the original background, branched over the codimension-2 boundary of the subregion in question. This is a singular manifold with a $2\pi n$ conical excess angle, and thus the associated partition function of a continuum quantum field theory is divergent. Moreover, it is sensitive to short distance physics near the entangling surface.

In contrast, the Euclidean gravity path integral instructs us to compute the Renyi partition function by integrating over metrics that are smooth (more precisely, the Einstein-Hilbert action dynamically suppresses metrics with conical singularities). The result is that the Renyi partition function can be evaluated in perturbation theory around a gravity saddle which is completely smooth in the interior. In generic situations without time reflection symmetry, this will be a complex saddle of the Euclidean path integral \cite{Rangamanietal}. The result is finite and insensitive to short distance physics. 

An important point is that the relation between the Euclidean path integral and Hilbert space formulations of gravity is very subtle. In particular, the Euclidean path integrals associated to the Renyi partition functions do not have any obvious interpretation as a trace over a Hilbert space, precisely because the Euclidean time replica circle shrinks in the interior. This is closely connected with the fact that the Hilbert space obtained by quantizing gravity in connected spacetimes with two asymptotic boundaries does not factorize, as was shown in \cite{FactorizationProblem} for JT gravity.

Recall that even in quantum field theory, the Hilbert space does not factorize into pieces associated to spatial subregions. This is due to the fact that all states in the full Hilbert space have entanglement between degrees of freedom that are arbitrarily close to the entangling surface. This is the Lorentzian realization of the fact that the Renyi entropies are ill-defined partition functions on manifolds with conical singularities.

\begin{figure}
	\centering
	\includegraphics[width=0.9\linewidth]{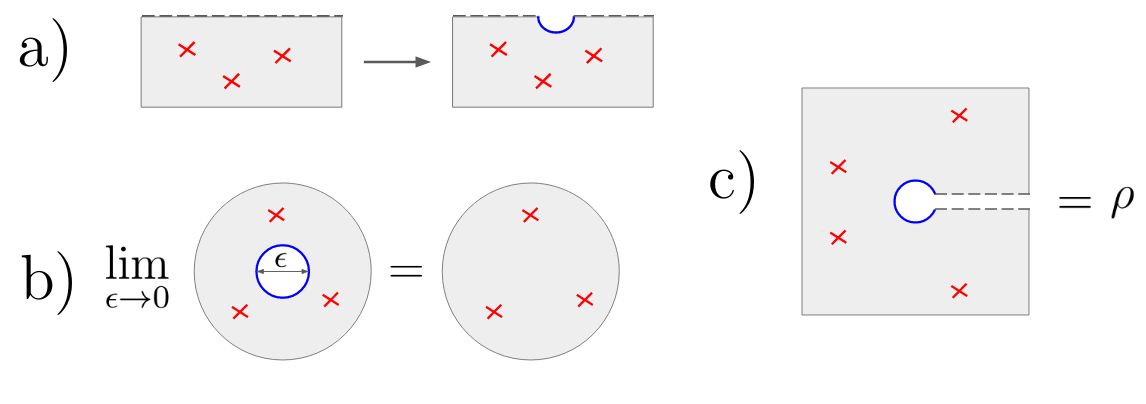}
	\caption{We illustrate the brick wall method of computing entanglement entropy in quantum field theory. In a), we show how a brick wall can be used to factorize an arbitrary state, prepared with the Euclidean path integral with the possible insertion of some local operators. The condition that the factorization map is an isometry in the limit $\epsilon \rightarrow 0$ is visualized in b). In c), we show how the brick wall may be used to construct the (unnormalized) reduced density matrix of a state. The Renyi partition functions are then given by $\text{Tr } \rho^n$. On the dashed lines, we fix the values of the fields that appear in the path integral. The brick wall is colored blue.}
	\label{fig:figure}
\end{figure}

The Renyi partition function can be regularized by excising the conical singularity with a codimension-1 boundary.\footnote{Sometimes Renyi entropies have been regularized by smoothening the replica manifold. However this does not lead to a Hilbert space interpretation.} This is directly related to a factorization map of the Hilbert space, $\calI_\epsilon: {\cal H} \rightarrow {\cal H}_L \otimes {\cal H}_R$. See Figure \ref{fig:figure}. The parameter $\epsilon$ specifies the size of the excised region. Each tensor factor is defined as the Hilbert space of the quantum field theory in the region with a boundary condition, placed at a distance $\epsilon$ from the original entangling surface. The resulting regularization of the entanglement entropy is called a brick wall regulator.

The limit of removing the regulator, $\epsilon \rightarrow 0$, is divergent, and the quantities of interest in quantum field theory are  derivatives with respect to various parameters, for example the location of the entangling surface, which may remain finite. In general it is a very nontrivial question whether the result depends on the choice of boundary condition.

In some respects then, the gravitational case is simpler, because the microscopic entanglement entropy itself is finite. Ideally one would like to find a local boundary condition in gravity that defines an isometric cutting map $\calI: {\cal H} \rightarrow {\cal H}_L \otimes {\cal H}_R $ via the Euclidean path integral. If this were possible, then the Renyi entropies computed with respect to the factorization defined by $\calI$ would equal the Renyi entropies computed from the replica trick using Euclidean path integral. The resulting entanglement entropy would match the full holographic answer.

However, in gravity, defining an isometric cutting map with a local boundary condition on a brick wall is impossible for a very general reason. The essential point is that in the gravitational path integral with boundary, there is no way to impose that the extension of the metric into the excised region is smooth, (that is, that the conical angle is $2\pi$) because this condition is non-local in replica time. Equivalently, one needs to reproduce terms in the bulk action that involve integrals of the Ricci scalar over the excised region (as these are what suppress conical singularities there). In 2d gravity, the Gauss-Bonnet theorem relates this to an integral of the extrinsic curvature $\frac{1}{2\pi}\int \sqrt{\gamma} \kappa - 1 $. Because of the 1 which measures the Euler character of the disk, this cannot be written as a local integral on the boundary. The problem is topological in nature and its analog appears in gravity in any dimension.

Similarly, the Euclidean Schwarzian description of JT gravity \cite{stanfordwitten} includes the constraint that the field that appears in the path integral is globally a diffeomorphism of the circle. That is, it has winding 1. This is again a non-local condition.

 Furthermore, in the first order formulation of gravity, the same issue presents itself as a topological restriction on the spin connection that cannot be imposed by a local boundary condition.

Therefore there is no local boundary condition in gravity for which the cutting map defined by the path integral is an isometry. Instead the Lorentzian interpretation of the geometric entropy is slightly different.

Recall from the previous subsection that the Euclidean JT path integral with \eqref{eq:introbc} imposed on a brick wall is equivalent to the Euclidean Schwarzian theory \cite{stanfordwitten}, which we discuss in Section \ref{sec:euclidpathhh}. To compute the disk path integral, one must compactify Euclidean time: $\tau \sim \tau + \beta$. The action contains a single field, $\phi(\tau)$, which is required to satisfy the nonlocal condition that $\phi(\tau  + \beta) = \phi(\tau) + 2\pi$. In Section \ref{sec:euclidpathhh}, we explain how the disk path integral can be given a Hilbert space interpretation. First, we integrate in new fields and write the action in a canonical form. The field $\phi(\tau)$ becomes a noncompact scalar. The Euclidean disk path integral can be computed, up to normalization,\footnote{The normalization is off by a factor of infinity. This infinity can be thought of as the extremal entropy. We are not interested in such contributions.} as
\begin{equation}
\label{eq:windingone}
Z_{\text{Disk}} \propto \text{Tr } D_E \, e^{-\beta H},
\end{equation}
where
\begin{equation}
\label{eq:windingone2}
D_E = e^{- 2 \pi i \, \pi_\phi},
\end{equation}
and $\pi_\phi$ is canonically conjugate to $\phi$. The role of $D_E$ is to enforce the winding constraint. Usually, to compute the partition function of a compact scalar in quantum mechanics, we must sum over all winding sectors. If we are only interested in the sector with winding number 1, we can decompactify the scalar and use \eqref{eq:windingone} and \eqref{eq:windingone2} to isolate the desired winding sector.

The field $\phi$ represents an angular coordinate on the hyperbolic disk. The winding number constraint is simply the condition that $\phi$ has periodicity $2 \pi$, or that there is no conical singularity in the hyperbolic disk. Thus, while it is not possible to use a local boundary condition to ensure that the conical angle around the brick wall in the disk path integral is $2 \pi$, this condition can easily be enforced by imposing \eqref{eq:introbc} on the brick wall and inserting the operator $D_E$ into the path integral.

Equation \eqref{eq:windingone} provides a Hilbert space interpretation of the disk path integral in the Euclidean Schwarzian theory. However, we want a Hilbert space interpretation in the Lorentzian theory. The key feature of the Euclidean computation above is that we can use \eqref{eq:introbc} to define a cutting map, and the reduced density matrix of the Hartle-Hawking state that is produced by this cutting map is $e^{-\beta H}$. However, the norm of the Hartle-Hawking state is given by $\text{Tr } D_E e^{-\beta H}$ instead of $\text{Tr } e^{-\beta H}$. Furthermore, the $n$th Renyi partition function is given by $\text{Tr } D_E e^{- \beta n H}$ instead of $\text{Tr } e^{-\beta n H}$. The usual formula for the Renyi partition function is modified by inserting an extra universal operator that fixes the conical angle around the brick wall to be $2 \pi$.

The lesson we learn from the Euclidean Schwarzian path integral is that the formula in the Lorentzian theory for the $n$th Renyi partition function for \emph{any} state should be given by $Z[n] = \text{Tr } D \tilde{\rho}^n$, where $\tilde{\rho}$ is a density matrix produced by a cutting map.\footnote{The matrix $\tilde{\rho}$ may not be normalized because the cutting map is not an isometry. We define $\tilde{\rho}$ in \eqref{eq:rhotilde}.} We will show in Section \ref{sec:splitting} that $D$ is Hermitian and that it commutes with the Hamiltonian and with the reduced density matrix $\tilde{\rho}$ for any state of the original unfactorized Hilbert space. Let $\calI : \calh \rightarrow \calh_L \otimes \calh_R$ denote the cutting map defined with \eqref{eq:introbc} on the brick wall.\footnote{Because the Euclidean and Lorentzian theories have different contours as mentioned earlier, computing $\calI$ directly from the Euclidean path integral is subtle. We describe below how we can explicitly determine $\calI$.} The Hilbert space of the unfactorized Lorentzian JT theory is given by $\calh$, while $\calh_L \otimes \calh_R$ corresponds to two copies of the Lorentzian Schwarzian theory. If $\tilde{\rho}$ is the reduced density matrix of the state $\ket{\Psi} \in \calh$ produced by $\calI$, then $\tilde{\rho}$ must satisfy
\begin{equation}
\label{eq:rhotilde}
\tilde{\rho} = \text{Tr}_{\calh_L} \, \calI \ket{\Psi} \bra{\Psi} \calI^\dagger. 
\end{equation}
While $\calI$ is not an isometry, our Euclidean calculations imply that there exists an operator $D$ that acts on $\calh_R$ only such that
\begin{equation}
\braket{\Psi | \Psi} = \text{Tr}_{\calh_R} D \tilde{\rho}.
\end{equation}
If we define the factorization map $\calJ : \calh \rightarrow \calh_L \otimes \calh_R$ to satisfy\footnote{The square root of an operator is not uniquely defined. However, only the precise definition of $D$ will matter for us.}
\begin{equation}
\calJ = \sqrt{D} \calI,
\end{equation}
then $\calJ$ is an isometry. 

We mentioned earlier that given $\calh_L \otimes \calh_R$ in JT theory, there is only one physically sensible isometric factorization map into this factorized Hilbert space. This factorization map, which we describe in detail in Section \ref{sec:splitting}, is the only possible candidate for $\calJ$. With the explicit formula for $\calJ$ in hand, we can determine explicit formulas for $\calI$ and $D$. To do this, we use the principle that the reduced density matrix produced by $\calI$ for the Hartle-Hawking state must be proportional to $e^{- \beta H}$, or the thermal density matrix of the Lorentzian Schwarzian theory. This must be the case because $\calI$ is a cutting map defined with a local, covariant, boundary condition on the brick wall. Thus, to determine $D$, we compute the reduced density matrix of the Hartle-Hawking state using $\calJ$, and we compare the result with $e^{-\beta H}$. The formula for $D$ is given in \eqref{eq:defectoperatordefinition}. The formula for $\calI$ then follows from the formulas for $\calJ$ and $D$. In what follows, we only need the formulas for $\calJ$ and $D$.

To check the statement that $\calJ$ results from composing a simple, universal, operator and a cutting map produced by the Euclidean path integral, we explicitly show in Section \ref{sec:euclidpathhh} how we use the Euclidean Schwarzian path integral with an insertion of $\sqrt{D_E} = e^{- i \pi \, \pi_\phi}$ to compute the image of the Hartle-Hawking state under $\calJ$. Upon analytically continuing the expression for the path integral in the appropriate way, we obtain the factorized Hartle-Hawking wavefunction in $\calh_L \otimes \calh_R$. In Appendix \ref{sec:euclidpathintegral}, we also compute the reduced density matrix of the Hartle-Hawking wavefunction from the Euclidean Schwarzian path integral with an insertion of $D_E$. In this case, the analytic continuation is more subtle, but the results are consistent with equation \eqref{eq:rhofmtrixelement}, which is a formula for the reduced density matrix of an arbitrary state in the image of $\calJ$.

With explicit formulas for $\calJ$ and $D$ in hand, we can compute the Renyi partition functions for any state. The $n$th Renyi partition function is given by
\begin{equation}
\label{eq:introzndefect}
Z[n] = \text{Tr } D \tilde{\rho}^n = \text{Tr } D^{1 - n} \rho^n, 
\end{equation}
where 
\begin{equation}
\rho \equiv D \tilde{\rho}.
\end{equation}
The reduced density matrix $\rho$ is produced by the factorization map $\calJ$ and it satisfies $\text{Tr } \rho = 1$ because $\calJ$ is an isometry. Starting from \eqref{eq:introzndefect}, we can compute the entanglement entropy as follows:
\begin{equation}
\label{eq:introee}
\left.-\partial_n \left(\frac{Z[n]}{Z[1]^n}\right)\right|_{n = 1} = \text{Tr } \left[(\log D) \rho \right] - \text{Tr } \rho \log \rho.
\end{equation}
In Section \ref{sec:ee}, we explicitly check that the first term on the right hand side of \eqref{eq:introee} exactly matches the expectation value of the area operator in \eqref{eq:sgen}. That is, in JT gravity, we can define two separate notions of the area term in the quantum HRT formula. The first definition is simply the first term on the right hand side of \eqref{eq:introee}. The second definition is the expectation value of the area operator in JT gravity, which can be unambiguously computed in the unfactorized Hilbert space. It is nontrivial to check that these two definitions are equivalent in JT gravity. We expect that our logic involving the defect operator can be generalized to higher dimensions, where it is not as obvious how to precisely define the area operator.

The defect operator is exactly what would be produced in the long distance effective theory by performing the trace over the microscopic degrees of freedom in the black hole microstates, starting with the exact factorization of the full theory. The universality of this result in the long distance theory is the fact that this merely implements the appropriate rotation in the Euclidean path integral, given by $D_E$.

\section{JT gravity}
\label{sec:JTgravity}
In this section we briefly review basic aspects of JT gravity. For more detail, see \cite{FactorizationProblem}. We quantize Lorentzian JT gravity on the spatial interval. Using the notation of \cite{FactorizationProblem}, the action is given by
\begin{equation} S_{JT} = \Phi_0 \left(\int_M d^2 x \sqrt{-g}R + 2 \int_{\partial M}dt  \sqrt{|\gamma|}K\right) + \int_M d^2 x \sqrt{-g} \Phi (R + 2) + 2 \int_{\partial M} dt \sqrt{|\gamma|} \Phi(K-1),
\label{eq:action} \end{equation}
where we have set $16 \pi G_N = 1$. On both boundaries, we impose the boundary conditions
\begin{equation}
\label{eq:boundaryconditions}
\gamma_{tt} = -\frac{1}{\epsilon^2}, \quad \Phi = \frac{\phi_b}{\epsilon},
\end{equation}
and we work in the limit $\epsilon \rightarrow 0$.

The classical solutions describe two-sided black holes. The phase space is parameterized by the horizon value of the dilaton $\Phi_h$, which is restricted to be positive, and the relative time shift between the two boundaries. The first term in \eqref{eq:action} is purely topological and determines the extremal entropy of the black hole solutions. The semiclassical entropy of the black hole is given by
\begin{equation}
\label{eq:entropy}
S = 4 \pi (\Phi_0 + \Phi_h).
\end{equation} The black hole energy measured from the left boundary equals the energy measured from the right boundary, and the total Hamiltonian is given by
\begin{equation}
\label{eq:hamiltonian}
H_{\text{tot}} = H_L + H_R = \frac{2 \Phi_h^2}{\phi_b}.
\end{equation}

To quantize the theory, it is convenient to integrate out $\Phi$ in the bulk so that the off-shell path integral amounts to integrating over metrics with constant negative curvature that satisfy \eqref{eq:boundaryconditions} on the boundaries. We parameterize each off-shell geometry as a cutout of global AdS$_2$, with metric
\begin{equation}
\label{eq:globalads2}
ds^2 = \frac{-dT^2 + d\sigma^2}{\sin^2 \sigma}, \quad \sigma \in (0,\pi).
\end{equation}
Let $\sigma = 0$ correspond to the right boundary of global AdS$_2$ and $\sigma = \pi$ correspond to the left boundary. The right cutout boundary is defined by the trajectory $(T_R(t),\sigma_R(t))$, where $T_R(t)$ can be regarded as a reparameterization of the boundary time $t$. The boundary conditions \eqref{eq:boundaryconditions} imply that
\begin{equation}
\label{eq:boundarypositionofsigmar}
\sin \sigma_R = \epsilon \dot{T}_R + \mathcal{O}(\epsilon^3).
\end{equation}
Likewise, the location of the left boundary is determined by $T_L(t)$. In the limit $\epsilon \rightarrow 0$, \eqref{eq:action} becomes
\begin{equation}
\label{eq:actionwithschwarzians}
S_{JT} =  \phi_b \int dt \left[- \dot{T}_R^2 + \left(\frac{\ddot{T}_R}{\dot{T}_R}\right)^2\right] + \phi_b \int dt \left[- \dot{T}_L^2 + \left(\frac{\ddot{T}_L}{\dot{T}_L}\right)^2\right],
\end{equation}
where we have ignored the topological term and additional unimportant total derivative terms. The JT theory with two boundaries simplifies to two copies of the Schwarzian theory. Each Schwarzian theory has a global $\widetilde{SL}(2,\mathbb{R})$ symmetry\footnote{By $\widetilde{SL}(2,\mathbb{R})$ we mean the universal cover of $SL(2,\mathbb{R})$.} that corresponds to applying an isometry of AdS$_2$ to the reparameterization field. We must gauge the subgroup of $\widetilde{SL}(2,\mathbb{R}) \times \widetilde{SL}(2,\mathbb{R})$ transformations that corresponds to applying the same isometry to both $T_R$ and $T_L$, since these transformations do not change the cutout geometry.\footnote{In our conventions, this is not the diagonal \sltr subgroup. We provide more details in Section \ref{sec:splitting}.} Furthermore, note that a Cauchy surface at time $t$ ends on the left boundary at time $T_L(t)$ and on the right boundary at time $T_R(t)$. Because Cauchy surfaces are spacelike, physical wavefunctions can only have support for $|T_L - T_R| < \pi$. These constraints also appeared in \cite{kitaevsuh}.

To summarize, the physical states of JT gravity can be described as states of two copies of the Schwarzian theory, subject to a gauged \sltr symmetry and the constraint that the wavefunction vanishes for $|T_L - T_R| > \pi$. In the next section, we discuss the Schwarzian theory in more detail.

\section{Schwarzian theory}
\label{sec:schwarziantheory}
In this section, we quantize the Schwarzian theory. As described above, the action of the Schwarzian theory is
\begin{equation}
\label{eq:schwarzianaction}
S = \phi_b \int dt \left[- \dot{T}^2 + \left(\frac{\ddot{T}}{\dot{T}}\right)^2\right].
\end{equation}
Because $T(t)$ is a reparameterization field, its derivative $\dot{T}(t)$ is constrained to be positive. We rewrite the action by integrating in three new fields:
\begin{equation}
\label{eq:firstorderaction}
 S = \int dt \left[p_T \dot{T} + p_\chi \dot{\chi} - \frac{1}{2 \phi_b} \left( \frac{p_\chi^2}{2} + p_T e^\chi + \frac{e^{2 \chi}}{2}\right) \right]. \end{equation}
The path integral is defined with a flat measure for all of the fields:
\begin{equation}
\label{eq:pathintegral}
\int \mathcal{D}T \, \mathcal{D}\chi \, \mathcal{D}p_T \, \mathcal{D}p_\chi \, e^{i S}.
\end{equation}
Upon integrating out $p_T$, $\chi$, and $p_\chi$, we recover \eqref{eq:schwarzianaction} with a reparameterization-invariant measure for $T$. The quantization of \eqref{eq:firstorderaction} is straightforward because we have written the action in a canonical form. For the remainder of this section, we let $T$, $p_T$, $\chi$, and $p_\chi$ denote the quantum operators that obey the usual canonical commutation relations. The Hamiltonian obeys\footnote{This Hamiltonian should be thought of as either $H_L$ or $H_R$ in \eqref{eq:hamiltonian}.}
\begin{equation}
\label{eq:schwarzianhamiltonian}
2 \phi_b H =   \frac{p_\chi^2}{2} + p_T e^\chi + \frac{e^{2 \chi}}{2}.
\end{equation}
The three $\widetilde{SL}(2,\mathbb{R})$ generators are
\begin{align}
\label{eq:generators}
J_1 &= p_T, \\
J_2 &= (\cos T) \, p_T - (\sin T) \, p_\chi + e^\chi \cos T + \frac{i \sin T}{2},  \\
J_3 &= (\sin T) \, p_T + (\cos T) \, p_\chi + e^\chi \sin T - \frac{i \cos T}{2},
\end{align}
which satisfy the following commutation relation:
\begin{equation}
\label{eq:commutator}
[J_a,J_b] = i \epsilon\indices{_a_b^c} J_c,
\end{equation}
where $\epsilon_{abc}$ is the completely antisymmetric tensor  satisfying $\epsilon_{123} = 1$, and indices are raised and lowered with $\eta_{ab} = \text{diag}(-1,1,1)$. The infinitesimal \sltr transformations of $T$ and $\chi$ are
\begin{align}
\label{eq:infinitesimaltrans}
\begin{split}
i [J_1,T] = 1, \quad i [J_2,T] = \cos T, \quad i[J_3,T] = \sin T, \\
i [J_1,\chi] = 0, \quad i [J_2,\chi] = - \sin T, \quad i[J_3,\chi] = \cos T.
\end{split}
\end{align}
The quadratic Casimir of the algebra is related to the Hamiltonian as follows:
\begin{equation}
\label{eq:casimir} \frac{1}{2}\left(\eta^{ab} J_a J_b - \frac{1}{4}\right) = \frac{p_\chi^2}{2} + p_T e^{\chi} + \frac{1}{2} e^{2 \chi} = 2 \phi_b H. \end{equation}
The central element of $\widetilde{SL}(2,\mathbb{R})$ is $e^{2 \pi i \, p_T}$. Representations of $\widetilde{SL}(2,\mathbb{R})$ are labeled by the eigenvalues of the Casimir and the central element \cite{SL2RNotes}.

We choose to work in a basis that diagonalizes $H$ and $p_T$. For a fixed eigenvalue of $p_T$, Schrodinger's equation for $\chi$ is equivalent to the quantum mechanics of a particle moving in the Morse potential \cite{MorsePotential}. For fixed $p_T > 0$, the Hamiltonian in \eqref{eq:schwarzianhamiltonian} has a positive spectrum, corresponding to scattering states. For fixed $p_T < 0$, the potential develops a well, allowing for negative energy bound states. Because this well becomes arbitrarily deep for sufficiently negative values of $p_T$, the Hamiltonian in \eqref{eq:schwarzianhamiltonian} is unbounded from below.

We only consider the positive energy states, as the negative energy states are not relevant for our later analysis. For $s >0$ and $m \in \mathbb{R}$, let $\ket{s \, m}$ denote a basis state that satisfies
\begin{align}
\label{eq:energyquantumnumber}
2 \phi_b H \ket{s \, m} &= \frac{s^2}{2} \ket{s \, m}, \\
p_T \ket{s \, m} &= m \ket{s \, m}.
\end{align}
Comparing \eqref{eq:energyquantumnumber} with \eqref{eq:hamiltonian}, we see that $s$ equals twice the horizon value of the dilaton. The wavefunction in the $T,\chi$ basis that corresponds to $\ket{s \, m}$ is
\begin{equation}
\label{eq:whittakerwavefunction}
\braket{T \, \chi | s \, m} = \frac{e^{i m T}}{\sqrt{2 \pi}}\frac{\sqrt{s \sinh 2 \pi s}}{\sqrt{2} \pi} |\Gamma(\frac{1}{2} + m + i s)| e^{- \frac{\chi}{2}} W_{-m,i s}( 2 e^\chi ), \end{equation}
where $W_{a,b}(x)$ is the Whittaker W-function. The state $\ket{s \, m}$ is normalized so that
\begin{equation}
\braket{s \, m | s^\prime \, m^\prime} = \delta(s - s^\prime)\delta(m - m^\prime).
\end{equation}
In the $T,\chi$ basis, the projection operator onto a single positive-energy representation is given by
\begin{equation} \label{eq:projector}\begin{gathered} P_{s,\mu}(T_2,\chi_2,T_1,\chi_1) \equiv \sum_{m \in \mathbb{Z} + \mu} \braket{T_2 \, \chi_2| s \, m} \braket{s \, m| T_1 \, \chi_1} = \\
\frac{i e^{2 \pi i \mu \floor*{\frac{T_2 - T_1}{2\pi}}}}{4 \pi}
	\frac{s \sinh (2 \pi  s)}{\cosh (2 \pi  s)+\cos (2 \pi  \mu )}  \left| \csc \left(\frac{T_2-T_1}{2}\right) \right| e^{i \left(e^{\chi_1}+e^{\chi_2}\right) \cot\left(\frac{T_2-T_1}{2}\right)} \\
	 \times \left( e^{-\pi  s+2  \pi i \mu } H_{2 i s}^{(1)}\left[2 e^{\frac{\chi_1+\chi_2}{2}} \left|\csc \left(\frac{T_2-T_1}{2}\right)\right|\right]- e^{\pi  s} H_{2 i s}^{(2)}\left[2 e^{\frac{\chi_1+\chi_2}{2}} \left| \csc \left(\frac{T_2-T_1}{2}\right) \right| \right]\right), \end{gathered}\end{equation}
where $\mu \in [0,1)$ determines the eigenvalue of the central element. We explain how we derived this expression in Appendix \ref{sec:solveschwarzian}. The factor of $e^{2 \pi i \mu \floor*{\frac{T_2 - T_1}{2\pi}}}$ ensures that
\begin{equation} \label{eq:pperiodicity}
P_{s,\mu}(T_2 + 2 \pi n,\chi_2,T_1,\chi_1) = e^{2 \pi i \mu n }P_{s,\mu}(T_2,\chi_2,T_1,\chi_2), \quad n \in \mathbb{Z}.
\end{equation}
Finally, we note that when $T_2 = T_1$, the projector becomes
\begin{equation}
\label{eq:sett2equalst1}
P_{s,\mu}(T_1,\chi_2,T_1,\chi_1) = \frac{e^{- \chi_1}}{2 \pi} \frac{s \sinh 2 \pi s}{\cosh (2 \pi s) + \cos (2 \pi \mu)}  \delta(\chi_1 - \chi_2).
\end{equation}
The trace of the projector is proportional to infinity. This infinity is the origin of the infinite additive constant in the entanglement entropies that we compute. For now, we define the infinite constant $V$ so that
\begin{equation}
\label{eq:defofv}
\int_{-\infty}^\infty d T \int_{-\infty}^{\infty} d \chi \, P_{s,\mu}(T,\chi,T,\chi) =  \frac{s \sinh 2 \pi s}{\cosh (2 \pi s) + \cos (2 \pi \mu)}  V.
\end{equation}
It is important that $V$ does not depend on either $s$ or $\mu$. As explained in Appendix \ref{sec:solveschwarzian}, we should think of $V$ as the volume of \sltr.

\section{Factorizing the Hilbert space}
\label{sec:splitting}

Defining entanglement entropy in quantum field theory is subtle because the Hilbert space cannot be decomposed into factors that correspond to different spatial subregions. The Hilbert space of JT gravity is also not factorizable, as explained in \cite{FactorizationProblem}. A standard approach to defining entanglement entropy is to isometrically embed the physical Hilbert space into a larger, factorized Hilbert space that contains unphysical states. For instance, to compute the entanglement entropy of an interval in 2D Yang-Mills, one can map the Hilbert space to a larger space that includes states that are not gauge-invariant at the endpoints of the interval \cite{yangmills2d}. As explained at the end of Section \ref{sec:JTgravity}, the Hilbert space of JT gravity consists of two copies of the Hilbert space of the Schwarzian theory subject to the constraints that physical states are invariant under a gauged \sltr global symmetry and that Cauchy surfaces are spacelike. To isometrically map this Hilbert space to a factorized Hilbert space, we simply remove these constraints. This factorization map follows naturally from describing JT gravity as a system of two particles propagating near the two boundaries of AdS$_2$. In turn, the boundary particle description follows naturally from a local boundary condition placed on a brick wall in the Euclidean path integral, as we explain in Section \ref{sec:defect}. We further explain in Section \ref{sec:euclidpathhh} how we construct this factorization map using the Euclidean path integral as in Figure \ref{fig:cuttingfigure}, with the extra insertion of a ``half-defect.'' In this section, we describe the details of our factorization map (which we denote by $\calJ$ throughout this paper) and compute the reduced density matrix of an arbitrary state in JT gravity. Then, we compute the reduced density matrix of the Hartle-Hawking state and observe that it does not take the form of a thermal density matrix. Rather, it is given by a thermal density matrix times a defect operator that does not depend on the temperature.

As reviewed in \cite{symmetriesnearhorizon}, the isometries of AdS$_2$ consist of global time translations, spatial translations, and boosts. The associated Killing vector fields are
\begin{align}
\label{eq:kvf}
B &= -\cos T \, \cos \sigma \partial_T + \sin T \, \sin \sigma \partial_\sigma, \\
P &= \sin T \, \cos \sigma \partial_T + \cos T \, \sin \sigma \partial_\sigma, \\
E &= \partial_T.
\end{align}
Using \eqref{eq:boundarypositionofsigmar} and a corresponding relation for $T_L$, we find that $T_R$ and $T_L$ transform under $B$, $P$, and $E$ as follows:
\begin{align}
\label{eq:BPEtrans}
\begin{split}
\delta_B T_R = - \cos T_R, \quad \delta_P T_R = \sin T_R, \quad  \delta_E T_R = 1, \\
\delta_B T_L = \cos T_L, \quad \delta_P T_L = -\sin T_L, \quad  \delta_E T_L = 1.
\end{split}
\end{align}
Let $J_a^L$ and $J_a^R$ correspond to the generators given in \eqref{eq:generators} for the left and right Schwarzian theories, respectively. Let $\Psi(T_L,\chi_L,T_R,\chi_R)$ be any state in the Hilbert space of the two Schwarzian theories. Using \eqref{eq:infinitesimaltrans} and \eqref{eq:BPEtrans}, we find that if $\Psi$ is to be invariant under the isometries of global AdS$_2$, $\Psi$ must satisfy
\begin{equation}
\label{eq:conditions}
(J_1^L + J_1^R) \Psi = 0, \quad (J_2^L - J_2^R) \Psi = 0, \quad (J_3^L - J_3^R) \Psi = 0.
\end{equation}
In addition to imposing \eqref{eq:conditions}, we also impose the condition that $\Psi$ vanish for $|T_L - T_R| > \pi$, which ensures that the left and right cutout boundaries are always spacelike separated. These conditions fix $\Psi$ completely up to a single function of the renormalized geodesic distance \cite{FactorizationProblem} between the two boundaries:
\begin{equation}
\label{eq:mostgeneralpsi}
\Psi(T_L,\chi_L,T_R,\chi_R) = \left\{ \begin{array}{cc}
e^{-\frac{1}{2}(\chi_L + \chi_R)} e^{i(e^{\chi_R} - e^{\chi_L})\tan(\frac{T_L-T_R}{2})} F\left(\frac{e^{\chi_R + \chi_L}}{\cos^2 \left(\frac{T_R -T_L}{2}\right)}\right), & |T_L - T_R| < \pi, \\
0, & |T_L - T_R| \ge \pi.
\end{array}  \right. \end{equation}

To complete our definition of the factorization map, we must find gauge-invariant solutions to Schrodinger's equation. Let $\Psi_s$ denote a wavefunction that satisfies \eqref{eq:mostgeneralpsi} for $F = F_s$. If we impose that the energy measured from the left is $H_L = \frac{s^2}{4 \phi_b}$ for $s > 0$, then Schrodinger's equation becomes
\begin{equation}
\label{eq:schrodinger}
2 \phi_b H_L \ket{\Psi_s} = \frac{s^2}{2} \ket{\Psi_s} = \frac{1}{2}\left(\eta^{ab} J^L_a J^L_b - \frac{1}{4}\right) \ket{\Psi_s}.
\end{equation}
This implies that $F_s(u)$ is a linear combination of $\sqrt{u} \,  K_{2 i s}(2 \sqrt{u})$ and $\sqrt{u} \, I_{2 i s}(2 \sqrt{u})$. In order for $\Psi_s(T_L,\chi_L,T_R,\chi_R)$ to vanish as $|T_L - T_R|$ approaches $\pi$, $F_s(u)$ must be proportional to $\sqrt{u} \, K_{2is}(2 \sqrt{u})$. Note that gauge invariance implies that the energy of $\Psi_s$ measured from the right is the same as the energy measured from the left, because as shown in \eqref{eq:schwarzianhamiltonian}, the energy is determined by the Casimir. Also, there are no gauge-invariant wavefunctions that are energy eigenstates with negative energy. The constraint $(J^L_1 + J^R_1) \Psi = 0$ is incompatible with the fact that the Hamiltonian in \eqref{eq:schwarzianhamiltonian} only has negative energy eigenstates in the $p_T < 0$ sector.

The factorization map is defined to map an energy eigenstate in JT gravity of total energy $H_{\text{tot}} = \frac{s^2}{2 \phi_b}$ to the unique gauge-invariant state of two copies of the Schwarzian theory with the same total energy, which is
\begin{equation} \begin{split}
&\Psi_s(T_L,\chi_L,T_R,\chi_R) \\
 &= \left\{
\begin{array}{cc}
\frac{\sqrt{2 s \sinh(2 \pi s)}}{2\pi^2} \frac{e^{i(e^{\chi_R} - e^{\chi_L})\tan(\frac{T_L-T_R}{2})}}{\cos\left(\frac{T_R - T_L}{2}\right)} K_{2 i s}\left(\frac{2 e^{\frac{\chi_L + \chi_R}{2}}}{ \cos\left(\frac{T_R - T_L}{2}\right)}\right)  , &   |T_L - T_R| < \pi,
\\
0, & |T_L - T_R| > \pi.
\end{array}
\right.
\label{eq:imageoffactorizationmap} \end{split}\end{equation}
The normalization of \eqref{eq:imageoffactorizationmap} is chosen so that
\begin{equation}
\label{eq:norm}
\braket{\Psi_{s^\prime} | \Psi_s} = V \delta(s - s^\prime),
\end{equation}
where $V$ was defined in \eqref{eq:defofv}.

Next, we present the reduced density matrix that one obtains after tracing over the Hilbert space of the left Schwarzian theory. For any function $f(s)$ that is square-integrable on $\mathbb{R}^+$, define
\begin{equation}
\label{eq:psif}
\ket{\Psi_f} \equiv \int_0^\infty ds \, f(s) \ket{\Psi_s},
\end{equation}
with $\ket{\Psi_s}$ given in \eqref{eq:imageoffactorizationmap}. Without loss of generality, we assume throughout this paper that the function $f(s)$ always satisfies
\begin{equation}
\int_0^\infty ds \, |f(s)|^2 = 1.
\end{equation}
We find that the reduced density matrix of $\ket{\Psi_f}$ is given by
\begin{equation}
\label{eq:rhof}
\rho_f \equiv \frac{\text{Tr}_L \ket{\Psi_f}\bra{\Psi_f}}{\braket{\Psi_f | \Psi_f}} = \frac{1}{V}\int_0^\infty ds \, |f(s)|^2 \int_{-\infty}^\infty dm \frac{\cosh 2 \pi s + \cos 2 \pi m}{s \sinh 2 \pi s} \ket{s \, m} \bra{s \, m}. \end{equation}
Using \eqref{eq:projector}, we find that in the $T,\chi$ basis, $\rho_f$ is given by
\begin{equation}
\label{eq:rhofmtrixelement}
\begin{split} \braket{T_2 \, \chi_2 | \rho_f | T_1 \, \chi_1} = \frac{1}{V}\int_0^\infty ds \, |f(s)|^2 \left(\frac{i e^{i(e^{\chi_1} + e^{\chi_2})\cot\left(\frac{T_2 - T_1}{2}\right)}}{4 \pi} \left| \csc\left(\frac{T_2 - T_1}{2}\right)\right|\right) \\ \times
\left\{  \begin{array}{c c}
-e^{\pi s} H^{(2)}_{2is}\left(2 e^{\frac{\chi_1 + \chi_2}{2}} \csc\left(\frac{T_2 - T_1}{2}\right)\right), & 0 < T_2 - T_1 < 2 \pi,  \\
e^{- \pi s} H^{(1)}_{2 i s}\left(2 e^{\frac{\chi_1 + \chi_2}{2}} \left|\csc\left(\frac{T_2 - T_1}{2}\right)\right|\right), & - 2\pi < T_2 - T_1 < 0,\\
0, & |T_2 - T_1| > 2 \pi.
\end{array}  \right. \end{split} \end{equation}
We derive these expressions in Appendix \ref{sec:solveschwarzian}.

The Hartle-Hawking state was computed exactly in various bases in \cite{zhenbinyang} by performing the Euclidean path integral of JT gravity on the semicircle. The image of the Hartle-Hawking state under the factorization map is
\begin{equation}
\label{eq:hhstate}
\ket{\Psi_\beta} \equiv \int_0^\infty ds \, \sqrt{2 s \sinh 2 \pi s} \, e^{- \beta \frac{s^2}{4 \phi_b} }\ket{\Psi_s},
\end{equation}
with $\ket{\Psi_s}$ given in \eqref{eq:imageoffactorizationmap}. We choose the overall normalization of \eqref{eq:hhstate} to agree with our calculation of the Hartle-Hawking wavefunction in Section \ref{sec:euclidpathhh}, which uses the Euclidean path integral.
The explicit wavefunction, which follows from \eqref{eq:imageoffactorizationmap}, is
\begin{equation} \begin{split}
&\braket{T_L \, \chi_L \, T_R \, \chi_R|\Psi_\beta} \\
&= \left\{
\begin{array}{cc}
\int_0^\infty ds\frac{  \, s \sinh(2 \pi s)}{\pi^2 \cos\left(\frac{T_R - T_L}{2}\right)} e^{-\beta \frac{s^2}{4 \phi_b}} e^{i(e^{\chi_R} - e^{\chi_L})\tan(\frac{T_L-T_R}{2})} K_{2 i s}\left(\frac{2 e^{\frac{\chi_L + \chi_R}{2}}}{\cos\left(\frac{T_R - T_L}{2}\right)} \right)  , &   |T_L - T_R| < \pi,
\\
0, & |T_L - T_R| > \pi.
\end{array}
\right.
\label{eq:HHwavefunction} \end{split}\end{equation}
Note that $\ket{\Psi_\beta}$ satisfies
\begin{equation}
e^{- u H_L}\ket{\Psi_\beta} = e^{- u H_R}\ket{\Psi_\beta} = \ket{\Psi_{\beta + u}}.
\end{equation}
Furthermore, the norm $\braket{\Psi_{\frac{\beta}{2}}|\Psi_{\frac{\beta}{2}}}$ agrees (up to a $\beta$-independent, multiplicative infinite constant) with the Euclidean disk path integral of JT gravity \cite{stanfordwitten}. From \eqref{eq:rhof}, the reduced density matrix is
\begin{equation}
\label{eq:rhohh}
\rho_\beta \equiv \frac{\text{Tr}_L \ket{\Psi_\frac{\beta}{2}}\bra{\Psi_\frac{\beta}{2}}}{\braket{\Psi_\frac{\beta}{2}|\Psi_\frac{\beta}{2}}} = \frac{1}{V \, N_\beta } \int_0^\infty ds \int_{-\infty}^\infty dm   \quad   2 (\cosh 2 \pi s + \cos 2 \pi m) \, e^{- \beta \frac{s^2}{4 \phi_b}} \ket{s \, m} \bra{s \, m}, \end{equation}
where
\begin{equation}
\label{eq:nbeta}
 N_\beta = V^{-1} \braket{\Psi_\frac{\beta}{2}|\Psi_\frac{\beta}{2}} =  \int_0^\infty ds \, 2 s \sinh 2 \pi s \, e^{- \beta \frac{s^2}{4 \phi_b} }. 
\end{equation}
If we define the following operator,
\begin{equation}
\label{eq:defectoperatordefinition}
D \equiv 2 (\cosh 4 \pi \sqrt{ \phi_b H} + \cos 2 \pi p_T),
\end{equation}
which acts on the Hilbert space of the Schwarzian theory, we find from \eqref{eq:rhohh} that the Euclidean disk path integral of JT gravity, which is proportional to the norm of the Hartle-Hawking state, satisfies
\begin{equation}
\label{eq:partitionfunctionwithdefectoperator}
Z_{\text{Disk}}(\beta) \propto \braket{\Psi_{\frac{\beta}{2}}|\Psi_{\frac{\beta}{2}}} = \text{Tr } D e^{-\beta H},
\end{equation}
where the trace is over positive energy states of the Schwarzian theory. This is not the usual formula for the partition function of a quantum mechanical system. We refer to $D$ as the defect operator.\footnote{Another approach to quantizing JT gravity is given in \cite{exactquantization}, which makes use of a codimension-1 defect in the bulk. It would be interesting to relate this defect to our defect operator. See also \cite{kitaevsuh}, where the black hole partition function takes the form of \eqref{eq:partitionfunctionwithdefectoperator}.} In the Lorentzian theory, the defect operator $D$ is defined by comparing $\text{Tr}_L \ket{\Psi_\frac{\beta}{2}}\bra{\Psi_\frac{\beta}{2}}$ with $e^{-\beta H}$. Our rationale is that a factorization map defined from a local boundary condition on a brick wall in the Euclidean path integral would have mapped the Hartle-Hawking state to the thermofield double state. Thus, given the factorization map, we can determine the defect operator $D$ from the image of the Hartle-Hawking state.

In Sections \ref{sec:defect} and \ref{sec:euclidpathhh}, we explain the role of the defect operator in the Euclidean theory. In Section \ref{sec:ee}, we return to the Lorentzian theory where we use the defect operator to compute the entanglement entropy of any state. We will show that the contribution of the defect operator agrees with the expectation value of the area operator in the quantum HRT formula.

\section{Gravitational path integral with brick wall boundary}

\label{sec:defect}

The notion of entanglement entropy requires a factorization of the Hilbert space. In continuum quantum field theory, the full Hilbert space does not factorize according to division between spatial regions, due to UV modes near the edge of the region. This is reflected in the fact that the Renyi partition functions are on singular manifolds with conical excess angles, and so are ill-defined. More physically, the result depends on short distance physics near the conical singularity, and so is not a property of the long distance effective theory. In some cases, the dependence of the entanglement entropy on various parameters can be immune to such short distance physics, and it is that which is usually discussed.

In contrast, in the exact holographic theories of quantum gravity, the entanglement entropy, defined with respect to subregions of the boundary, is UV finite. The Euclidean gravity computation of the Renyi partition functions is over smooth bulk geometries, without any conical singularity (more precisely, the Einstein-Hilbert action suppresses the contribution of singularity geometries dynamically). For this reason the result, to leading approximation, is independent of the UV completion of the long distance gravity effective theory.

However in the long distance effective theory of gravity, the Hilbert space does not factorize. This is in some sense for the opposite reason as the case of continuum quantum field theories. Namely, in gravity, there are no nontrivial diffeomorphism invariant observables localized in a finite region, rather than too large an algebra of such operators. To understand the Lorentzian meaning of the gravitational Renyi calculation, one must define a cutting map from the gravitational Hilbert space to a factorized space. 

To define a cutting map, we impose a local boundary condition on a brick wall as shown in Figure \ref{fig:cuttingfigure}. In this section, we show that after imposing the appropriate boundary condition on the brick wall, the Euclidean path integral formally simplifies to the path integral of \cite{stanfordwitten}. We also point out that a local boundary condition cannot fix the conical angle around the brick wall to be $2 \pi$. 

Because we are working with the Euclidean theory, our conventions are slightly different from the rest of the paper. We will write the Euclidean action in first-order variables. Define the three sl(2) generators $T_a$ as follows:
\begin{equation}
T_1 = \frac{\sigma_1}{2}, \quad T_2 = \frac{\sigma_3}{2}, \quad  T_3 = \frac{i \sigma_2}{2},
\end{equation}
where $\sigma_a$ refers to the Pauli matrices. We can package the frame one-forms $e^1,e^2$ and the spin connection one-form $\omega$ into a gauge field as follows:
\begin{equation}
A = e^1 T_1 + e^2 T_2 + \omega T_3.
\end{equation}
The dilaton is given by $\Phi$. The first-order formalism requires two additional scalar fields, $\lambda^1$ and $\lambda^2$. We can package them into the adjoint scalar field $B$ as follows:
\begin{equation}
B = \lambda^1 T_1 + \lambda^2 T_2 + \Phi T_3.
\end{equation}
We define the path integral on a spatial interval with coordinate $x$ times Euclidean time $\tau$, which has period $\beta$. Setting $16 \pi G = 1$, the Euclidean action is given by
\begin{equation}
\label{eq:Euclideanfirstorderaction}
I =  4\int_{\text{bulk}} \text{Tr } B F + 2\int_{\text{AdS}} \left[\hat{\Phi} \omega - \delta_{ab} \lambda^a (e^b - \hat{e}^b) \right] + 2\int_{\text{wall}} \left[\Phi \omega - \delta_{ab} \lambda^a e^b \right],\end{equation}
where $F = dA + A \wedge A$, $\delta_{ab} = \text{diag}(1,1)$, and $\hat{e}^a$ and $\hat{\Phi}$ represent fixed background fields. The first term is the bulk action, the second term is the action for the asymptotic AdS boundary, and the third term is a boundary term on the brick wall. On the AdS boundary, we fix the metric and the dilaton:
\begin{equation}
e_\tau^a = \hat{e}_\tau^a, \quad \Phi = \hat{\Phi},
\end{equation}
while on the brick wall, we impose
\begin{equation}
\label{eq:brickwallbc}
A_\tau = 0.
\end{equation}
The boundary terms in \eqref{eq:Euclideanfirstorderaction} are chosen such that these constraints arise as classical equations of motion.

To simplify the Euclidean path integral, we integrate $A_\tau$ in the bulk along an imaginary contour so that it acts as a Lagrange multiplier enforcing the condition $\partial_x B = [B,A_x]$. We also introduce a new group-valued field $g \in \widetilde{SL}(2,\mathbb{R})$ such that $A_x = g^{-1} \partial_x g$. Next, we define
\begin{equation}
h \equiv g_{\text{wall}}^{-1} \, g_{\text{AdS}},
\end{equation}
which represents a Wilson line of the flat connection $A$ between the brick wall and the AdS boundary. The Euclidean action then simplifies to
\begin{equation}
\label{eq:simplifiedEuclideanfirstorderaction}
I = 2\int_0^\beta d\tau \left[ 2 \text{Tr } \left[B \, h^{-1} \partial_\tau h\right] - \delta_{ab} \lambda^a \hat{e}_\tau^b\right],
\end{equation}
where $B$ above is the value of the adjoint scalar on the AdS boundary. To define the path integral measure, we use the Haar measure for $h$ and a flat measure for $B$. We parameterize $h \in \widetilde{SL}(2,\mathbb{R})$ using global AdS$_3$ coordinates as follows:
\begin{equation}
\label{eq:parameterizeh}
h = e^{\phi T_3} e^{\rho T_1} e^{v T_3}, \quad \phi \in \mathbb{R}, v \in \mathbb{R}, \rho > 0,
\end{equation}
with the following identification:
\begin{equation}
(\phi,v) \sim (\phi + 2 \pi, v - 2\pi).
\end{equation}
We integrate $\lambda^1$ and $\lambda^2$ along an imaginary contour to impose the constraints
\begin{equation}
\begin{split}
\label{eq:constraints}
(\partial_\tau {\rho}, \partial_\tau {\phi} \, \sinh {\rho}) \cdot (\cos {v},\sin {v}) &= \hat{e}^1_\tau \\
(\partial_\tau {\rho}, \partial_\tau {\phi} \, \sinh {\rho}) \cdot (-\sin {v},\cos {v}) &= \hat{e}^2_\tau.
\end{split}
\end{equation}
These constraints imply that
\begin{equation}
(\partial_\tau \rho)^2 + (\sinh \rho \, \partial_\tau \phi)^2 = \delta_{ab} \hat{e}^a_\tau \hat{e}^b_\tau,
\end{equation}
meaning that the path integral simplifies to an integral over paths in a hyperbolic space with metric
\begin{equation}
\label{eq:hyperbolicdisk}
ds^2 = d\rho^2 + \sinh^2 \rho \, d\phi^2,
\end{equation}
where the coordinate $\phi$ is noncompact. To obtain the Schwarzian theory, we set $\hat{\Phi}$ and $\hat{e}^a_\tau$ as follows:
\begin{equation}
\hat{\Phi} = \frac{\phi_b}{\epsilon}, \quad \hat{e}^1_\tau = \frac{1}{\epsilon}, \quad \hat{e}^2_\tau = 0.
\end{equation}
Solving the constraints in the limit $\epsilon \rightarrow 0$, we find that
\begin{equation}
\label{eq:firstorderextrinsiccurvature}
\epsilon (\partial_\tau \phi ) \cosh \rho = 1 + \frac{1}{2}\left((\partial_\tau \phi)^2 - \frac{(\partial_\tau^2 \phi)^2}{(\partial_\tau \phi)^2}\right) \epsilon^2 + \mathcal{O}(\epsilon^4).
\end{equation}
Using \eqref{eq:firstorderextrinsiccurvature}, the action \eqref{eq:simplifiedEuclideanfirstorderaction} simplifies to
\begin{equation}
I = 2\phi_b\int_0^\beta d\tau \left[ \frac{1}{2}\left(-(\partial_\tau \phi)^2 + \frac{(\partial_\tau^2 \phi)^2}{(\partial_\tau \phi)^2}\right) \right],
\end{equation}
where we have ignored total derivative terms and canceled a divergent $\frac{1}{\epsilon^2}$ term with a counterterm.

The purpose of these manipulations is to show that  the Euclidean Schwarzian path integral of \cite{stanfordwitten} can be interpreted in terms of the JT gravity disk path integral with a brick wall. As mentioned in \eqref{eq:brickwallbc}, the brick wall boundary condition sets the frame fields and the spin connection to zero. This creates a cusp singularity in the bulk geometry. To obtain a smooth geometry around the brick wall, we instead want to impose the conditions
\begin{equation}
\label{eq:smoothbrickbc}
e^a_\tau = 0, \quad \int_{\text{wall}} d\tau \, \omega_\tau = 2 \pi.
\end{equation}
Note that the above condition for the spin connection, which ensures that the conical angle around the brick wall is $2\pi$, is nonlocal in $\tau$ and thus it is not possible to cut a path integral with this boundary condition to obtain a density matrix (as in Figure \ref{fig:figure}c). Equation \eqref{eq:smoothbrickbc} fixes the holonomy of the flat connection around the disk to be the central element of \sltr. From \eqref{eq:parameterizeh}, this is equivalent to constraining the field $\phi(\tau)$ to obey
\begin{equation}
\phi(\tau + \beta) = \phi(\tau) + 2 \pi.
\end{equation}
This is the winding constraint of \cite{stanfordwitten}.\footnote{Off-shell configurations where $v$ has nonzero winding number are suppressed in the limit $\epsilon \rightarrow 0$. These correspond to cutout trajectories of the hyperbolic disk that self-intersect.} Because $\phi(\tau)$ is noncompact, the winding constraint is easily implemented by inserting the defect operator $D_E$, which shifts the value of $\phi$ by $2 \pi$.

In this section, we showed how a local boundary condition on a brick wall in the Euclidean JT path integral leads us to the Euclidean Schwarzian path integral. We have also pointed out the need to insert the defect operator $D_E$ into the partition function of the Euclidean Schwarzian theory. In the next section, we provide more explicit details and we use the path integral to map the Hartle-Hawking wavefunction to the wavefunction in two copies of the Lorentzian Schwarzian theory.

\section{Computing the factorized Hartle-Hawking wavefunction with the Euclidean path integral}

\label{sec:euclidpathhh}

In this section, we explain how the Euclidean Schwarzian path integral can be used to prepare the factorized Hartle-Hawking state. Our strategy is similar to that of \cite{kitaevsuh}. We begin with the Euclidean disk path integral computed in \cite{stanfordwitten} (see also \cite{MSY}), which we reproduce below, using slightly different conventions:
\begin{equation}
\label{eq:swpathintegral}
Z_{\text{Disk}} = \int \frac{d\mu[\phi]}{\text{SL}(2,\mathbb{R})} \exp\left(-2 \phi_b \int_0^\beta d\tau \, \left(  \frac{1}{2} \left(\frac{\phi^{\prime \prime}}{\phi^\prime}\right)^2  - \frac{1}{2} ( \phi^\prime)^2  \right)\right).
\end{equation}
The field $\phi(\tau)$ is constrained to obey $\phi(\tau + \beta) = \phi(\tau) + 2 \pi$. This path integral computes the partition function of Euclidean JT gravity on the disk. The length of the disk is fixed and controlled by $\beta$. We have omitted the contribution from the topological term in the JT gravity action. The field $\phi(\tau)$ can be thought of as a reparameterization map from a circle of length $\beta$ to a circle of length $2\pi$. In particular, $\phi(\tau)$ defines a cutout geometry in the hyperbolic metric
\begin{equation}
\label{eq:hyperbolicmetric}
ds^2 = d\rho^2 + \sinh^2 \rho \, d\phi^2.
\end{equation}
The path integral in \eqref{eq:swpathintegral} gauges the SL(2,$\mathbb{R}$) transformations of $\phi(\tau)$ that produce equivalent cutout geometries. We choose to not divide by the infinite volume factor. This results in the addition of an infinite constant to the Hartle-Hawking entropy that we are not interested in. Rather, we are interested in how the entropy changes with temperature. Indeed, the infinite volume may be absorbed into the definition of the extremal entropy. Next, we integrate in additional fields, resulting in the following path integral:
\begin{equation}
\label{eq:introEuclideanschwarzianpathintegral}
\begin{split}
&K_\beta(\phi_2,\psi_2,\phi_1,\psi_1) \equiv
\\
&\int \mathcal{D}\phi \mathcal{D}\psi \mathcal{D}\pi_\phi \mathcal{D}\pi_\psi  \exp\left(\int_0^\beta d\tau \left[ i \pi_\psi \psi^\prime +  i \pi_\phi \phi^\prime - \frac{1}{2 \phi_b} \left[ \frac{\pi_\psi^2}{2} + i \pi_\phi e^\psi - \frac{1}{2} e^{2 \psi}\right]   \right]\right).
\end{split}
\end{equation}
At time $\tau = 0$ the fields $\phi,\psi$ are fixed to $\phi_1,\psi_1$ and at time $\tau = \beta$ the fields are fixed to $\phi_2,\psi_2$. To get \eqref{eq:swpathintegral} (without dividing by the volume of SL(2,$\mathbb{R}$)), one simply needs to set $\psi_1 = \psi_2 = \psi$ and $\phi_2 = \phi_1 + 2 \pi = \phi$ and integrate over $\phi$ and $\psi$:
\begin{equation}
\label{eq:zdiskfromtrace}
Z_{\text{Disk}} \cdot \text{Vol SL}(2,\mathbb{R}) = \int_{-\infty}^\infty d\psi \int_{0}^{2 \pi} d\phi \, K_\beta(\phi + 2 \pi,\psi,\phi,\psi).
\end{equation}
From now on, we no longer keep track of prefactors multiplying $Z_{\text{Disk}}$. The action in \eqref{eq:introEuclideanschwarzianpathintegral} is written in a canonical form. We let $\phi$ and $\psi$ be canonical position variables that are both valued in $(-\infty,\infty)$. The fields $\pi_\phi$ and $\pi_\psi$ are the canonical momenta. Thus, it is appropriate to write
\begin{equation}
\label{eq:euclidpathintegralbraket}
K_\beta(\phi_2,\psi_2,\phi_1,\psi_1) = \braket{\phi_2 \, \psi_2 | e^{- \beta H} | \phi_1 \, \psi_1 },
\end{equation}
with
\begin{equation}
H = \frac{1}{2 \phi_b} \left[ \frac{\pi_\psi^2}{2} + i \pi_\phi e^\psi - \frac{1}{2} e^{2 \psi}\right].
\end{equation}
Since this Hamiltonian is not Hermitian, computing \eqref{eq:euclidpathintegralbraket} is subtle. However, for $2 \pi > \phi_2 - \phi_1 > 0$, \eqref{eq:euclidpathintegralbraket} has been computed in \cite{zhenbinyang}. The result is\footnote{See Appendix \ref{sec:euclidpathintegral} for more details.}
\begin{equation}
\label{eq:introEuclideanschwarzianpropagator}
\begin{split}
K_\beta(\phi_2,\psi_2,\phi_1,\psi_1) = \exp\left((e^{\psi_2} + e^{\psi_1})\cot\left(\frac{\phi_1 - \phi_2}{2}\right) \right) \\
\times  \frac{1}{\pi^2} \frac{1}{\sin \frac{\phi_2 - \phi_1}{2}} \int_{0}^{\infty}ds s \sinh(2 \pi s) e^{-\frac{s^2}{4 \phi_b}\beta} K_{2is}\left(\frac{ 2 e^{\frac{\psi_2 + \psi_1}{2}}}{\sin \frac{\phi_2 - \phi_1}{2}}\right).
\end{split}
\end{equation}
If we define the Euclidean defect operator to be
\begin{equation}
\label{eq:defectoperatordefinitioneuclidean}
D_E \equiv e^{-2 \pi i \, \pi_\phi },
\end{equation}
then from \eqref{eq:zdiskfromtrace}, the disk partition function is given by
\begin{equation}
\label{eq:writezdiskastrace}
Z_{\text{Disk}} \propto \text{Tr } D_E e^{- \beta H}.
\end{equation}
The right hand side above is divergent, but this is expected because we chose to not divide the disk path integral by the volume of SL(2,$\mathbb{R}$).

\begin{figure}
	\centering
	\includegraphics[width=0.7\linewidth]{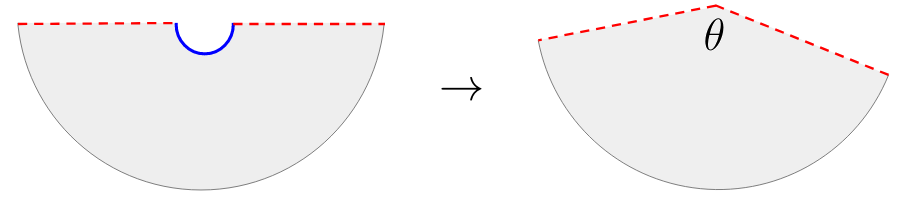}
	\caption{We illustrate the process of using the Euclidean path integral to compute the Hartle-Hawking wavefunction in a factorized Hilbert space. If the integral of the spin connection on the brick wall boundary (given in blue) satisfies $\int \omega = \theta$, then the geometry will have an opening angle of $\theta$ between the two boundaries that represent the two Hilbert space factors. For $\theta \neq \pi$, there is a kink. An isometric factorization map cannot have a kink, or else the norm of the Hartle-Hawking wavefunction in the factorized Hilbert space would be incorrect.}
	\label{fig:kinkfigure}
\end{figure}

To gain a better intuition for how the Euclidean Schwarzian path integral may be used to define a cutting map, recall the procedure for defining a cutting map, given in Figure \ref{fig:cuttingfigure}. We impose boundary conditions on all boundaries and integrate over all possible geometries. Because we impose in \eqref{eq:introbc} that the metric component along the brick wall vanishes, the brick wall shrinks to a point in the geometry defined by the metric. The spin connection on the brick wall determines the opening angle between the two boundaries that meet at the brick wall. See Figure \ref{fig:kinkfigure}. To define an isometric factorization map, we want the spin connection on the brick wall to satisfy $\int \omega = \pi$. However, this cannot be enforced by a local boundary condition on the brick wall. Because we impose the local boundary condition \eqref{eq:introbc} on the brick wall, we must insert $\sqrt{D_E} = e^{- i \pi \, \pi_\phi}$ into the Euclidean Schwarzian path integral.

The factorization map defined by inserting $\sqrt{D_E}$ into the Euclidean Schwarzian path integral is exactly the same factorization map $\calJ$ described in Section \ref{sec:splitting}. We now explicitly demonstrate this for the Hartle-Hawking state. The Euclidean propagator \eqref{eq:introEuclideanschwarzianpropagator} with the insertion of a ``half-defect'' is defined to be
\begin{equation}
\label{eq:propwithhalfdefect}
\begin{split}
K_\beta(\phi_2 + \pi,\psi_2,\phi_1,\psi_1) &= \frac{1}{\pi^2} \int_0^\infty ds \, e^{- \beta \frac{s^2}{4 \phi_b}} \, s \sinh(2 \pi s)  \frac{e^{(e^{\psi_1} + e^{\psi_2})\tan \frac{\phi_2 - \phi_1}{2}}}{\cos \frac{\phi_2 - \phi_1}{2}} K_{2 i s}\left(\frac{2 e^{\frac{\psi_1 + \psi_2}{2}}}{\cos \frac{\phi_2 - \phi_1}{2}}\right).
\end{split}
\end{equation}
Adding $\pi$ to $\phi_2$ on the left-hand-side can also be intuitively thought of as evolving halfway around the thermal circle. To get the Hartle-Hawking wavefunction in the Lorentzian theory, we analytically continue $\phi_2 \rightarrow i \phi_R$ and $\phi_1 \rightarrow -i \phi_L$.\footnote{The coordinates $\phi_R$ and $\phi_L$ are Rindler time coordinates in the right and left Rindler wedges, respectively. Increasing values of $\phi_R$ and $\phi_L$ both correspond to time going forwards in the global AdS$_2$ geometry that the Rindler wedges are embedded in. This explains the sign difference between $\phi_2 \rightarrow i \phi_R$ and $\phi_1 \rightarrow - i \phi_L$.} The result is
\begin{equation}
\label{eq:propwithhalfdefectanalyticallycontinued}
K_\beta(i \phi_R + \pi,\psi_2,-i \phi_L,\psi_1) = \frac{1}{\pi^2} \int_0^\infty ds \, e^{- \beta \frac{s^2}{4 \phi_b}} \, s \sinh(2 \pi s)  \frac{e^{i(e^{\psi_L} + e^{\psi_R})\tanh \frac{\phi_L + \phi_R}{2}}}{\cosh \frac{\phi_L + \phi_R}{2}} K_{2 i s}\left(\frac{2 e^{\frac{\psi_1 + \psi_2}{2}}}{\cosh \frac{\phi_L + \phi_R}{2}}\right).\end{equation}
We can interpret \eqref{eq:propwithhalfdefectanalyticallycontinued} as the Hartle-Hawking wavefunction $\braket{\phi_L \, \psi_L \, \phi_R \, \psi_R | \Psi_\beta}$ given in Rindler variables, since upon analytically continuing $\phi$ in \eqref{eq:hyperbolicmetric}, one gets the metric for the Rindler wedge. Note that there is an ambiguity in the definition of the basis vectors $\ket{\phi \, \psi}$. We are free to perform a phase rotation for any function $f(\phi,\psi)$:
\begin{equation}
\ket{\phi \, \psi} \rightarrow e^{i f(\phi,\psi)}\ket{\phi \, \psi}.
\end{equation}
We choose the function $f(\phi,\psi)$ such that the final expression for the Hartle-Hawking wavefunction is
\begin{equation}
\label{eq:HHwavefunctionrindler}
\begin{split}
& \braket{\phi_L \, \psi_L \, \phi_R \, \psi_R|\Psi_\beta}
\\
= &\frac{1}{\pi^2} \int_0^\infty ds \, e^{- \beta \frac{s^2}{4 \phi_b}} \, s \sinh(2 \pi s)  \frac{e^{i(e^{\psi_L} + e^{\psi_R})\tanh \frac{\phi_L + \phi_R}{2}}}{\cosh \frac{\phi_L + \phi_R}{2}} K_{2 i s}\left(\frac{2 e^{\frac{\psi_L + \psi_R}{2}}}{\cosh \frac{\phi_L + \phi_R}{2}}\right)\times e^{-i e^{\psi_R} \tanh \phi_R - i e^{\psi_L} \tanh \phi_L}.
\end{split}
\end{equation}
To convert from Rindler time variables to global time variables, we use
\begin{equation}
\label{eq:rindlertoglobaltime}
\tan \frac{T_{L,R}}{2} = \tanh \frac{\phi_{L,R}}{2}, \quad e^{\chi_{L,R}} = e^{\psi_{L,R}} \text{sech } \phi_{L,R} , \quad d\chi_{L,R} \, dT_{L,R} = \frac{1}{\cosh \phi_{L,R}} \, d\phi_{L,R} \, d\psi_{L,R}. \end{equation}
The result is
\begin{equation}
\label{eq:waaveffunctioninpatch}
\braket{T_L \, \chi_L \, T_R \, \chi_R|\Psi_\beta} \\
=
\int_0^\infty ds\frac{  \, s \sinh(2 \pi s)}{\pi^2 \cos\left(\frac{T_R - T_L}{2}\right)} e^{-\beta \frac{s^2}{4 \phi_b}} e^{i(e^{\chi_R} - e^{\chi_L})\tan(\frac{T_L-T_R}{2})} K_{2 i s}\left(\frac{2 e^{\frac{\chi_L + \chi_R}{2}}}{\cos\left(\frac{T_R - T_L}{2}\right)} \right).
\end{equation}
Note that the Rindler variables only cover the patch of the global variables where $|T_L| < \frac{\pi}{2}$ and $|T_R| < \frac{\pi}{2}$. We want to determine the factorized Hartle-Hawking wavefunction in the Lorentzian JT theory for all values of the global time variables. Because \eqref{eq:waaveffunctioninpatch} only depends on $T_L$ and $T_R$ via their difference, we may analytically continue \eqref{eq:waaveffunctioninpatch} into the entire region where $|T_L - T_R| < \pi$. We next determine the wavefunction in the region $|T_L - T_R| > \pi$. Note that in Figure \ref{fig:kinkfigure}, one can fix the profile of the induced metric on the dashed red boundaries. The dashed red boundaries are always spacelike. Thus, the Euclidean path integral implies the constraint mentioned in Section \ref{sec:JTgravity} that the boundary particles are always spacelike separated. The wavefunction produced by the Euclidean path integral then agrees with \eqref{eq:HHwavefunction}. By considering the Hartle-Hawking wavefunction as an example, we have demonstrated how the Euclidean path integral with a half-defect computes the factorization map $\calJ$ that we described in Section \ref{sec:splitting}.

\section{Entanglement entropy}

\label{sec:ee}

In this section, we compute the entanglement entropy of arbitrary pure states in JT gravity. First, we compute the von Neumann entropy of reduced density matrices produced by our factorization map $\calJ$. We then argue that this von Neumann entropy reproduces $S_{\text{bulk}}$ in the quantum HRT formula. We explain how the expectation value of the area operator is defined in JT gravity. Finally, we use the defect operator to compute the Renyi partition functions. We check that the defect operator makes a universal contribution to the entanglement entropy that agrees with the expectation value of the area operator.

\subsection{von Neumann entropy of the reduced density matrix}

In this subsection, we compute the von Neumann entropy of reduced density matrices of states that are in the image of the factorization map defined in Section \ref{sec:splitting}. It follows from \eqref{eq:rhof} and \eqref{eq:defofv} that the $n$th Renyi partition function associated with $\rho_f$ is
\begin{equation}
\label{eq:renyioff}
Z_f[n] \equiv \text{Tr } \rho_f^n = V^{1-n}\int_0^\infty ds (|f(s)|^2)^n \int_0^1 d\mu \left[\frac{\cosh 2 \pi s + \cos 2 \pi \mu}{s \sinh 2 \pi s}\right]^{n-1} .
\end{equation}
To evaluate the $\mu$ integral, we use the result that for $n \in \mathbb{R}$,
\begin{equation}
\label{eq:legendrep}
\int_0^1 d \mu \left[\frac{\cosh 2 \pi s + \cos 2 \pi \mu}{\sinh 2 \pi s}\right]^n = P_n(\coth(2 \pi s)),
\end{equation}
where $P_n(z)$ is the $n$th Legendre polynomial.\footnote{This expression indicates how the Legendre polynomials may be analytically continued to $n \notin \mathbb{Z}_{\ge 0}$.} Thus,
\begin{equation}
Z_f[n] = V^{1-n} \int_0^\infty ds (|f(s)|^2)^n \frac{P_{n-1}(\coth 2 \pi s)}{s^{n-1}}.
\end{equation}
We need to use the fact that for $z > -1$, \cite{Szmytkowski}
\begin{equation}
\label{eq:derivlegendre}
\left. \partial_n P_n(z) \right|_{n = 0} =  \log \frac{z + 1}{2}.
\end{equation}
The von Neumann entropy of $\rho_f$ is then given by
\begin{align}
\label{eq:entropyofrhof}
\begin{split}
S_f &\equiv - \text{Tr } \rho_f \log \rho_f = \left. -\partial_n \left(\frac{Z_f[n]}{Z_f[1]^n}\right)\right|_{n = 1}, \\
&= \log V  -  \int_0^\infty ds \, |f(s)|^2 \left[\log \frac{|f(s)|^2}{2 s \sinh 2 \pi s} + 2 \pi s \right].
\end{split}
\end{align}

\subsection{Entanglement entropy of the Hartle-Hawking state}

\label{sec:entanglemententropyofhh}

To compute the entropy of the Hartle-Hawking state $\ket{\Psi_{\frac{\beta}{2}}}$, we set \begin{equation}|f(s)|^2 = \frac{2 s \sinh(2 \pi s) e^{- \beta \frac{s^2}{4 \phi_b}}}{N_\beta},\end{equation} which follows from \eqref{eq:hhstate}. We define $N_\beta$ in \eqref{eq:nbeta}. The von Neumann entropy of $\rho_\beta$ is then
\begin{equation}
\begin{split}
S_\beta &\equiv - \text{Tr } \rho_\beta \log \rho_\beta ,
\\
&= \log V + \log 2\int_0^\infty ds \,  s \sinh(2 \pi s) e^{- \beta \frac{s^2}{4 \phi_b}}
+ \frac{\int_0^\infty ds \,  s \sinh(2 \pi s) e^{- \beta \frac{s^2}{4 \phi_b}} \left[ \beta \frac{s^2}{4 \phi_b} - 2 \pi s \right]}{\int_0^\infty ds \,  s \sinh(2 \pi s) e^{- \beta \frac{s^2}{4 \phi_b}}} .
\end{split}
\end{equation}
In JT gravity, the value of the dilaton at the horizon $\Phi_h$ plays the role of the area operator. If we restore factors of $16 \pi G_N$, we see from \eqref{eq:entropy} that the entropy above extremality is
\begin{equation}
S - S_{\text{extremal}} = \frac{\Phi_h}{4 G_N}.
\end{equation}
Thus, we define the area operator of JT gravity to simply be $A \equiv \Phi_h$. We will continue to set $16 \pi G_N = 1$ as before, so that $\frac{A}{4 G_N} = 4 \pi A$. In terms of the total Hamiltonian \eqref{eq:hamiltonian}, $A$ is given by
\begin{equation}
\label{eq:areaoperator}
A= \sqrt{\frac{H_{\text{tot}} \phi_b}{2}}.
\end{equation}
The expectation value of $4 \pi A$ in the Hartle-Hawking state $\ket{\Psi_{\frac{\beta}{2}}}$ is
\begin{equation}
\label{eq:expectationvalueofareaoperator}
\braket{4 \pi A}_\beta \equiv
\frac{\braket{\Psi_{\frac{\beta}{2}}|4 \pi A|\Psi_{\frac{\beta}{2}}}}{\braket{\Psi_{\frac{\beta}{2}}|\Psi_{\frac{\beta}{2}}}} = \frac{ \int_0^\infty ds \, (2 \pi s)  s \sinh 2 \pi s \, e^{- \beta \frac{s^2}{4 \phi_b} }}{\int_0^\infty ds \,  s \sinh 2 \pi s \, e^{- \beta \frac{s^2}{4 \phi_b} }}.
\end{equation}
It follows that
\begin{equation}
\label{eq:finalentropyofhh}
\begin{split}
\braket{4 \pi A}_\beta + S_\beta &= \log V + \log 2\int_0^\infty ds \,  s \sinh(2 \pi s) e^{- \beta \frac{s^2}{4 \phi_b}}
+ \frac{\int_0^\infty ds \,  s \sinh(2 \pi s) \, \left[ \beta \frac{s^2}{4 \phi_b} \right] e^{- \beta \frac{s^2}{4 \phi_b}} }{\int_0^\infty ds \,  s \sinh(2 \pi s) e^{- \beta \frac{s^2}{4 \phi_b}}}, \\
&= \log \left[8 \pi^{3/2} V\right] + \frac{3}{2} + \frac{8 \pi^2 \phi_b}{\beta}   - \frac{3}{2} \log \frac{\beta}{\phi_b}.
\end{split}
\end{equation}
Up to an unimportant additive constant, this is the entropy of the two-sided black hole in JT gravity, given in \eqref{eq:introhartlehawkingentropy}. We have thus demonstrated how the black hole entropy may be computed in Lorentz signature in a way that agrees with the expectations of \cite{EW,FLM}.

The infinite constant in \eqref{eq:finalentropyofhh} appears because the Hilbert space of the Schwarzian theory is comprised of unitary representations of \sltr, which are infinite-dimensional. As shown in \eqref{eq:entropyofrhof}, $V$ is a universal constant that appears in the entanglement entropy of any state. Thus, the difference of the entropies of two different states is well-defined in JT gravity. The presence of the infinite constant in \eqref{eq:finalentropyofhh} signals the inability of the effective gravity theory to predict the extremal entropy of the black hole.

\subsection{Entanglement entropy of arbitrary pure states}

The entropy of the Hartle-Hawking state is known from the Euclidean path integral calculation, and we have shown how to reproduce it in a Lorentzian analysis. As shown in \eqref{eq:finalentropyofhh}, the entropy is a sum of the quantum-corrected area term and the von Neumann entropy of the reduced density matrix of the Hartle-Hawking state computed in the Schwarzian theory. The quantum HRT formula \cite{FLM} gives a natural proposal for computing the entanglement entropy of any state in JT gravity. We propose that the entanglement entropy of $\ket{\Psi_f}$ in \eqref{eq:psif} is simply the sum of the expectation value of $4 \pi A$ (with $A$ given in \eqref{eq:areaoperator}) and the von Neumann entropy \eqref{eq:entropyofrhof}. The entanglement entropy of $\ket{\Psi_f}$ is given by
\begin{equation}
\label{eq:generalentanglemententropy}
\braket{4 \pi A}_f + S_f = \log V  - \int_0^\infty ds \, |f(s)|^2 \left[\log |f(s)|^2 - \log (2 s \sinh 2 \pi s) \right].
\end{equation}
The right-hand-side of \eqref{eq:generalentanglemententropy} has appeared elsewhere in the literature \cite{JenniferLin}. In particular, this formula is related to the center variable method of computing entanglement entropy. In the center variable method, the image of the factorization map (which we denote by $\mathcal{H}$) is written as a direct sum of superselection sectors, where each superselection sector is a factorized Hilbert space:
\begin{equation}
\mathcal{H} = \bigoplus_\alpha \,  \mathcal{H}_L^\alpha \otimes \mathcal{H}_R^\alpha.
\end{equation}
We are interested in the case where
\begin{equation}
\dim \mathcal{H}^\alpha_L = \dim \mathcal{H}^\alpha_R \equiv \dim \alpha.
\end{equation}
The reduced density matrix $\rho$ must take the form
\begin{equation}
\label{eq:centerdecomposition}
\rho = \bigoplus_\alpha \, p_\alpha \, \rho_\alpha, \quad \sum_\alpha p_\alpha = 1.
\end{equation}
The coefficient $p_\alpha$ is the probability of measuring the state in the superselection sector labeled by $\alpha$. The von Neumann entropy of $\rho$ is then given by
\begin{equation}
\label{eq:centerentropy}
S(\rho) = -\sum_\alpha p_\alpha \log p_\alpha + \sum_\alpha p_\alpha S(\rho_\alpha), \quad S(\rho_\alpha) \equiv - \text{Tr } \rho_\alpha \log \rho_\alpha.
\end{equation}
Earlier computations of entanglement entropy in JT gravity \cite{finestructure} have treated JT gravity as a 2D BF theory. In 2D BF theory, where the superselection sectors correspond to representations of the gauge group, $S(\rho_\alpha)$ is given by
\begin{equation}
\label{eq:maxentropy}
S(\rho_\alpha) = \log \dim \alpha.
\end{equation}
To obtain agreement with \eqref{eq:generalentanglemententropy}, one could write down a continuum analog of \eqref{eq:centerentropy}:\footnote{Because the superselection sectors are labeled by a continuous parameter, the additive constant must be infinite.}
\begin{equation}
\label{eq:centerentropycontinuum}
S(\rho) = -\int_0^\infty ds \, p_s \log p_s + \int_0^\infty ds \, p_s \log (s \sinh ( 2 \pi s)) + \text{constant},
\end{equation}
where \eqref{eq:maxentropy} is replaced with the measure $s \sinh(2 \pi s)$. Note that \eqref{eq:centerentropycontinuum} agrees with \eqref{eq:generalentanglemententropy} for $p_s = |f(s)|^2$. Thus, our results are consistent with the earlier work of \cite{JenniferLin,finestructure}.

\subsection{Entanglement entropy from the defect operator}

By this point, we have shown how to apply the quantum HRT formula to JT theory. The area operator is simple to define, and the $S_{\text{bulk}}$ term can be computed with the factorization map described in Section \ref{sec:splitting}. However, we have not yet described how to compute the Renyi entropies. Furthermore, the point of this paper is to frame the computation of the entanglement entropy in a way that might generalize to higher-dimensional theories or theories with matter. Thus, we now explain a separate way of computing the entanglement entropy that highlights the contribution of the defect operator $D$.

We start with the Hartle-Hawking state. Recall from \eqref{eq:partitionfunctionwithdefectoperator} that we may express the disk amplitude of JT gravity as
\begin{equation}
Z_{\text{Disk}}(\beta) \propto \text{Tr } D e^{-\beta H} = V\int_0^\infty ds \,   2 s \sinh(2 \pi s) \, e^{- \beta \frac{s^2}{4 \phi_b}}, 
\end{equation}
where the trace is over positive-energy states of the Schwarzian theory. From the Euclidean path integral, we know that the $n$th Renyi partition function in gravity is defined (up to normalization) by
\begin{equation}
\label{eq:znb}
Z[n \beta] \equiv \text{Tr } D e^{- n \beta H}.
\end{equation}
We next write
\begin{equation}
\label{eq:znbrewrite}
\frac{Z[n \beta]}{Z[\beta]^n} = \text{Tr } D^{1-n} \rho_\beta^n.
\end{equation}
Note that $\rho_\beta$ is the reduced density matrix produced by $\calJ$, which is the factorization map that results from the Euclidean path integral with a half-defect inserted (as was shown in Section \ref{sec:euclidpathhh}). Thus, $\rho_\beta^n$ contains $n$ full defects. Alternatively, one can say that the Euclidean path integral that computes $\rho_\beta^n$ satisfies the condition $\int \omega = 2 \pi n$ on the brick wall. If we instead impose $\int \omega = 2\pi$ while leaving the length of the asymptotically-AdS boundary unchanged, the path integral computes the density matrix $\rho_{n \beta}$. By comparing $\rho_{\beta}^n$ with $\rho_{n \beta}$, we see that we must include the factor of $D^{1-n}$ in \eqref{eq:znbrewrite}. 

The full entanglement entropy of the Hartle-Hawking state is given by
\begin{equation}
\label{eq:hheedefect}
\left.-\partial_n \left(\frac{Z[n \beta]}{Z[\beta]^n}\right)\right|_{n = 1} = \text{Tr } \left[(\log D) \rho_\beta \right] - \text{Tr } \rho_\beta \log \rho_\beta.
\end{equation}
By construction, \eqref{eq:hheedefect} is the full entanglement entropy of the Hartle-Hawking state, equal to \eqref{eq:introhartlehawkingentropy} up to an irrelevant additive constant. We already showed in Section \ref{sec:entanglemententropyofhh} that the second term on the right-hand-side, $S_\beta$, is precisely equal to the $S_{\text{bulk}}$ term in the quantum HRT formula. Thus, the $\text{Tr } \left[(\log D) \rho_\beta \right]$ term must equal the area term in the HRT formula.

Now we turn to general states. Given $\rho_f$ in \eqref{eq:rhof}, we define
\begin{equation}
\label{eq:rewritezfntilde}
\tilde{Z}_f[n] \equiv \text{Tr } D^{1-n} \rho_f^n,
\end{equation}
which is not to be confused with $Z_f[n]$ in \eqref{eq:renyioff}. Our definition of $\tilde{Z}_f[n]$ is meant to mirror \eqref{eq:znbrewrite}. We must multiply $\rho_f^n$ by $D^{1-n}$ for the same reason that is given below \eqref{eq:znbrewrite}. We explicitly compute from \eqref{eq:rewritezfntilde}, \eqref{eq:rhof}, and \eqref{eq:defectoperatordefinition} that
\begin{equation}
\tilde{Z}_f[n] = V^{1-n} \int_0^\infty ds \, \frac{|f(s)|^{2n}}{(2 s \sinh 2 \pi s)^{n-1}}.
\end{equation} 
It follows that the $n$th Renyi entropy of $\ket{\Psi_f}$ is
\begin{equation}
S_n = \frac{1}{1 - n} \log \tilde{Z}_f[n].
\end{equation}
We may compute the full entanglement entropy of the state $\ket{\Psi_f}$ as
\begin{equation}
\label{eq:ztildefullee}
\begin{split}
\left.-\partial_n \left(\frac{\tilde{Z}_f[n]}{\tilde{Z}_f[1]^n}\right) \right|_{n = 1} = \log V  +  \int_0^\infty ds \left[ -|f(s)|^{2} \log |f(s)|^2 + |f(s)|^{2}\log (2 s \sinh 2 \pi s)\right].
\end{split}
\end{equation}
This is identical to \eqref{eq:generalentanglemententropy}, or the HRT formula. Starting from \eqref{eq:rewritezfntilde}, we also have that
\begin{equation}
\label{eq:entropyofgeneralstatedefectoperator}
\left.-\partial_n \left(\frac{\tilde{Z}_f[n]}{\tilde{Z}_f[1]^n}\right)\right|_{n = 1} = \text{Tr } \left[(\log D) \rho_f \right] - \text{Tr } \rho_f \log \rho_f.
\end{equation}
We showed in \eqref{eq:entropyofrhof} that the $- \text{Tr } \rho_f \log \rho_f$ term is simply $S_f$, or the $S_{\text{bulk}}$ term in the HRT formula. Thus, the $\text{Tr } \left[(\log D) \rho_f \right]$ term agrees with the area term. Note that while $D$ was defined by examining the Hartle-Hawking state, we can use $D$ to compute the entanglement entropy of any state in the way described above.

We have given two different definitions of the entanglement entropy of an arbitrary pure state in JT gravity, and we have checked that they are equivalent. The first definition, which uses the quantum HRT formula, is given in \eqref{eq:generalentanglemententropy}. The second definition uses \eqref{eq:rewritezfntilde} to define what we mean by the $n$th Renyi partition function for an arbitrary state. The entanglement entropy is then given in \eqref{eq:ztildefullee}. The benefit of the second definition is that it also provides a formula for the Renyi entropies.

\section{Discussion}

\label{sec:discussion}

The framework for computing quantum corrections to gravitational entanglement entropy was originally outlined in \cite{FLM} and expanded upon in \cite{EW,alliteration}. The $n$th gravitational Renyi entropy is unambiguously defined from a Euclidean path integral whose asymptotic boundary is an $n$-sheeted replica manifold. The authors of \cite{FLM,alliteration} give a natural prescription for analytically continuing in $n$, and they argue that the entanglement entropy can be organized as the sum of a quantum-corrected area term and a bulk entropy term. Our analysis uses a Euclidean path integral with a brick wall boundary to factorize the Hilbert space of the Lorentzian JT theory. Our prescription for factorizing the Hilbert space allows us to express the entanglement entropy of the Hartle-Hawking wavefunction in the same way as in \cite{FLM,alliteration}. We can use the quantum HRT formula to compute the entanglement entropy of any pure state in JT gravity, and our results are consistent with \cite{JenniferLin,finestructure}. With our factorization map, the reduced density matrix of the Hartle-Hawking state is, up to normalization, $e^{-\beta H}$ times a defect operator $D$, whose origin is the winding constraint of the Euclidean path integral performed in \cite{stanfordwitten}. The defect operator is needed to give a Lorentzian interpretation of the Renyi partition functions. For the Hartle-Hawking state, the $n$th Renyi partition function is $Z_\beta[n] = \text{Tr } D e^{- n \beta H}$. The defect operator makes a universal contribution to the entanglement entropy that nontrivially agrees with the area term in the HRT formula.

\subsection{The role of edge modes}

In the context of computing entanglement entropy in quantum field theory, the term ``edge modes'' refers to extra degrees of freedom that are introduced on the entangling surface in order to factorize the Hilbert space. In gauge theories, edge modes naturally arise from ungauging the gauge transformations that act nontrivially on the entangling surface. The edge modes correspond to representations of the symmetry group associated with the entangling surface. For specific examples, see \cite{JenniferLin2}. After introducing edge modes to factorize the Hilbert space, a generic reduced density matrix $\rho$ takes the form
\begin{equation}
\rho = \bigoplus_\alpha p_\alpha \, \rho_\alpha \otimes \left(\frac{I_\alpha}{\dim \alpha}\right),
\end{equation}
where $\alpha$ labels representations of the surface symmetry group. Each $\rho_\alpha$ corresponds to a state of the quantum fields away from the entangling surface, and the coefficients $p_\alpha$ sum to 1. The second tensor factor in each of the $\alpha$-blocks given above is the Hilbert space of the edge modes and thus has dimensionality $\dim \alpha$. The density matrix of the edge modes must be proportional to the identity operator $I_\alpha$ so that it is invariant under the action of the surface symmetry group. The entanglement entropy of $\rho$ then takes the form
\begin{equation}
S(\rho) = \sum_\alpha - p_\alpha \log p_\alpha - p_\alpha  (\text{Tr } \rho_\alpha \log \rho_\alpha) + p_\alpha \log \dim \alpha.
\label{eq:discussionentropy}
\end{equation}

It has been suggested \cite{JenniferLin2,DonnellyFridel} that upon treating gravity as a theory of gauged diffeomorphisms, the $\log \dim \alpha$ term may account for the universal Bekenstein-Hawking entropy. That is, the representation theory of the surface symmetry group should imply the existence of edge modes whose entanglement entropy across an entangling surface of area $\mathcal{A}$ could possibly reproduce the universal quantity $\mathcal{A}/4G_N$. Progress towards an explicit realization of this proposal in JT gravity has been made in \cite{finestructure,JenniferLin}. The authors of \cite{finestructure} describe JT gravity as a BF theory with a Plancherel measure that agrees with the inverse Laplace transform of the disk path integral. There, the entanglement entropy of the Hartle-Hawking state is computed as a von Neumann entropy, and the final expression takes the form of \eqref{eq:discussionentropy}, as pointed out in \cite{JenniferLin}.\footnote{The second term in the sum in \eqref{eq:discussionentropy} is zero because the theory has no propagating bulk degrees of freedom.} Furthermore, \cite{JenniferLin} identifies the Bekenstein-Hawking entropy \eqref{eq:semiclassicalentropy} with the $\log \dim \alpha$ term in the semiclassical limit. However, it is still unclear what the surface symmetry group is. The analysis of \cite{finestructure} seems to suggest that the surface symmetry group is SL(2,$\mathbb{R}$)$^+$, or the subset of SL(2,$\mathbb{R}$) matrices with all positive entries. However, SL(2,$\mathbb{R}$)$^+$ is not a group. It is a subsemigroup of SL(2,$\mathbb{R}$). It would be interesting if the representation theory of SL(2,$\mathbb{R}$)$^+$ described in \cite{finestructure} could somehow emerge from the formalism of \cite{DonnellyFridel} applied to JT gravity, despite the fact that \cite{DonnellyFridel} asserts that the physical surface symmetries form a group.

In our analysis, we treat JT gravity as a theory of boundary particles. The particles are only coupled through the constraints described in Section \ref{sec:JTgravity}. This formulation of JT gravity leads to a natural prescription for factorizing the Hilbert space. We simply lift the constraints. The von Neumann entropy of the reduced density matrix produced by this factorization map reproduces $S_{\text{bulk}}$ in \eqref{eq:sgen}. However, this von Neumann entropy, which is given in \eqref{eq:entropyofrhof}, does not take the form of \eqref{eq:discussionentropy}. Thus, it would be misleading to say that this entanglement entropy is due to edge modes that arise from ungauging a gauge symmetry at the entangling surface. Likewise, in our analysis, the area term in the HRT formula does not arise from counting edge modes. Instead, we use a defect operator to provide a Lorentzian interpretation of the Euclidean path integral that computes the Renyi partition functions. The role of the defect operator is to ensure that there is no conical singularity in the path integral.

Specifying the edge modes of a theory is essentially identical to specifying a factorization map. That is, in \eqref{eq:generalmap}, the label $m$ can be interpreted as counting the degeneracy of edge modes. The ``natural'' choice of edge modes depends upon the language used to formulate the theory. In \cite{finestructure}, the authors formulate JT gravity in a way that has direct parallels with BF theory. The edge modes of JT gravity are essentially the edge modes of the BF theory, and they account for the full black hole entropy. In our analysis, the ``edge modes'' that arise from removing the constraints do not account for the full black hole entropy. It would be interesting to identify the role of edge modes in the framework of \cite{exactquantization}, which makes use of a codimension-1 defect to compute the disk partition function.

Our main point is that the usual definition of edge modes (as extra degrees of freedom that arise from ungauging a gauge symmetry at the entangling surface) is ambiguous because a gauge symmetry is a redundancy in the description of the theory. In \cite{finestructure}, the gauge ``group'' is SL(2,$\mathbb{R}$)$^+$, while in our analysis, the group that we gauge consists of the \sltr transformations that translate both boundary particles in the same way. As mentioned in Section \ref{sec:mainresults}, our computations are based upon the principle that a local boundary condition should be used to factorize the Hilbert space \cite{Tachikawa}. In standard 2D BF theory, it happens to be the case that the ``natural'' choice of edge modes \cite{JenniferLin2} also arises by imposing the boundary condition that the boundary component of the gauge field vanishes.

\subsection{Future directions}

There are various directions that we wish to explore. First, we would like to generalize our entire analysis to JT gravity coupled to matter, which was studied in \cite{eternaltraversable,symmetriesnearhorizon}. Suppose that one had a  factorization map defined by a local boundary condition for the matter theory in a fixed spacetime background, together with appropriate counterterms given by local actions for the background metric \cite{DJ}. For example,  calculations of entanglement entropy in scalar theories have often been performed with a lattice regulator, and ending the lattice results in Neumann boundary conditions in the continuum limit. 

One can then split the full Hilbert space including dynamical gravity using that matter boundary condition together with the boundary condition we presented for JT gravity. The entanglement entropy computed after that factorization, together with the inclusion of the defect operator, should automatically agree with the entropy given by the Euclidean replica trick. 

An intriguing possibility for future exploration is whether one can find a model of induced JT gravity, in the sense of a model of 2d gravity and dilaton coupled to matter with no curvature terms in the action in the UV, and the usual JT Lagrangian generated by integrating out short distance matter modes. Clearly that would require a non-trivial coupling between the dilaton and the matter fields. If  such a model could be found, then in the UV description, no defect operator would be required, as there would be no suppression of conical defects by the action of the microscopic theory.

Given the results of \cite{JLMS}, it would be interesting to compute relative entropies of nearby states and compare with relative entropies of nearby states computed the SYK model. It would also be interesting to compute the entropy of bulk states that correspond to nonstatic spacetimes, which is relevant for \cite{pagecurve1}. Furthermore, in our Lorentzian analysis, we only computed the black hole entropy associated with the Euclidean path integral with trivial disk topology. It would be interesting to compute the higher genus corrections to the entropy in Lorentz signature. We would also like to generalize our analysis to the deformed Schwarzian theory introduced in \cite{TTbarQM}. 

Finally, we are interested in generalizing our computations to gravitational theories in three or more dimensions. In principle, by  Kaluza-Klein compactification this should reduce to a special case of 2d gravity coupled to a matter sector that includes all of the modes of the higher dimensional theory. One important feature of such a construction is that it should result in a version of the quantum HRT formula that is well-defined to all orders in $G_N$ perturbation theory. In particular, the bulk region and the analog of the area term will be determined by the local boundary condition and universal defect operator, defined in the factorized Hilbert space.

\section*{Acknowledgements} We would like to thank  Thomas Dumitrescu, Akash Goel, Lampros Lamprou, Juan Maldacena, Andy Strominger, and Edward Witten for stimulating and helpful discussions, and Ping Gao for collaboration in the initial stages of this project.  This work was supported in part by NSFCAREER grant PHY-1352084.

\appendix

\section{Solving the Schwarzian theory}
\label{sec:solveschwarzian}

In this section, we provide more details on how we solve the Schwarzian theory. It is useful to relate the Schwarzian theory to the theory of a nonrelativistic particle moving on the $\widetilde{SL}(2,\mathbb{R})$ manifold. First, we define $\widetilde{SL}(2,\mathbb{R})$ generators
\begin{equation}
\label{eq:ggenerators}
G_1 = \frac{1}{2}\left(\begin{array}{cc}
0 & 1 \\
-1 & 0
\end{array} \right), \quad G_2 = \frac{1}{2}\left(\begin{array}{cc}
0 & 1 \\
1 & 0
\end{array} \right), \quad G_3 = \frac{1}{2}\left(\begin{array}{cc}
1 & 0 \\
0 & -1
\end{array} \right).
\end{equation}
The generators satisfy
\begin{equation}
[G_a,G_b] = \epsilon\indices{_a_b^c} G_c,
\end{equation}
where $\epsilon_{abc}$ is the completely antisymmetric tensor that satisfies $\epsilon_{123} = 1$, and indices are raised and lowered with $\eta_{ab} = \text{diag}(-1,1,1)$. We parameterize a generic element of $\widetilde{SL}(2,\mathbb{R})$ as
\begin{equation}
\label{eq:parameterizationofg}
g = e^{T G_1} e^{\chi G_3} e^{\eta (G_2 - G_1)},
\end{equation}
with $T$, $\chi$, and $\eta$ all valued in $\mathbb{R}$. The following metric is invariant  under left and right group multiplication:
\begin{equation}
\label{eq:metric}
ds^2 = \text{Tr} \left[dg g^{-1} dg g^{-1}\right] = - \frac{1}{2} dT^2 + \frac{1}{2} d\chi^2 +  e^{-\chi} dT d\eta. \end{equation}
The corresponding Haar measure on $\widetilde{SL}(2,\mathbb{R})$ is
\begin{equation}
\label{eq:Haarmeasure}
\mathcal{D}g = e^{-\chi} \, dT d\chi d\eta.
\end{equation}
Comparing \eqref{eq:Haarmeasure} with \eqref{eq:defofv}, we see that $V$ is the volume of \sltr. The Hilbert space of the theory of a particle moving on $\widetilde{SL}(2,\mathbb{R})$ consists of functions on the group manifold that are square-integrable with respect to the Haar measure. The inner product is
\begin{equation}
\label{eq:F1F2norm}
\braket{F_1|F_2} = \int \mathcal{D}g \, F_1(g)^* F_2(g) =  \int dT d\chi d\eta \, e^{-\chi} \, F_1(T,\chi,\eta)^*F_2(T,\chi,\eta).
\end{equation}
To obtain the Hilbert space of the Schwarzian theory, we truncate the Hilbert space of a particle moving on $\widetilde{SL}(2,\mathbb{R})$ by restricting to functions $F$ that satisfy
\begin{equation}
\label{eq:restriction}
\partial_\eta F = i F.
\end{equation}
Given a function $F(T,\chi,\eta)$ that satisfies \eqref{eq:restriction}, it is convenient to define a new function $f(T,\chi)$ as follows:
\begin{equation}
\label{eq:littleffrombigf}
f(T,\chi) \equiv e^{- \frac{\chi}{2}} F(T,\chi,\eta)e^{-i\eta}.
\end{equation}
The inner product of two such functions $f_1(T,\chi)$ and $f_2(T,\chi)$ is given by
\begin{equation}
\label{eq:littlefinnerproduct}
\braket{f_2|f_1} = \int dT d\chi \, f_2^* (T,\chi) f_1(T,\chi),
\end{equation}
which is essentially the same inner product as \eqref{eq:F1F2norm} but without the $\eta$ integration.

Because $F(T,\chi,\eta)$ is defined on \sltr, one may define a \sltr transformation that acts on $F$ by left-multiplication. Given $h \in \widetilde{SL}(2,\mathbb{R})$, we define a new function $\tilde{F}$ as follows:
\begin{equation}
\label{eq:lefttransformation}
\tilde{F}(g) \equiv F(hg).
\end{equation}
We define $\tilde{f}$ as in \eqref{eq:littleffrombigf}:
\begin{equation}
\label{eq:littleftildefrombigftilde}
\tilde{f}(T,\chi) \equiv e^{- \frac{\chi}{2}} \tilde{F}(T,\chi,\eta)e^{-i\eta}.
\end{equation}
Given $f(T,\chi)$, one may construct $F$ using \eqref{eq:littleffrombigf}, transform $F$ to $\tilde{F}$ using \eqref{eq:lefttransformation}, and then construct $\tilde{f}$ using \eqref{eq:littleftildefrombigftilde}. This defines a group action of \sltr acting on $f(T,\chi)$. This group action leads to the generators in \eqref{eq:generators}. We think of $f(T,\chi)$ as an element of the Hilbert space of the Schwarzian theory, with inner product \eqref{eq:littlefinnerproduct}. The Casimir of the algebra leads to the Hamiltonian, as shown in \eqref{eq:casimir}.

Given this connection between the Schwarzian theory and the quantum mechanics of a particle moving on \sltr, it is clear that wavefunctions of the Schwarzian theory are determined by matrix elements of \sltr representations. Given a matrix element $F(T,\chi,\eta)$, one may construct the wavefunction $f(T,\chi)$ using \eqref{eq:littleffrombigf}. Matrix elements of \sltr representations are known in the literature. There are different sources available \cite{SL2RNotes, matrixelements}, but we have mostly used \cite{matrixelements}. While \cite{matrixelements} only contains matrix elements of $SL(2,\mathbb{R})$, \cite{matrixelements} contains enough information to deduce the matrix elements of \sltr that we need.

The wavefunction \eqref{eq:whittakerwavefunction} corresponds to a matrix element of \sltr in a mixed basis, which we may denote as $D^{s,\mu}_{m\rho}(g)$. The label $s$ denotes the Casimir and $\mu$ indicates that the eigenvalue of the central element is $e^{2 \pi i \mu}$. The value of the Casimir indicates that we are using principal series representations. The $m$ index corresponds to a basis that diagonalizes $G_1$, while the $\rho$ index corresponds to a basis that diagonalizes $G_2 - G_1$. The wavefunction \eqref{eq:whittakerwavefunction} then follows from (4.6) of \cite{matrixelements}, with the parameterization of the group element given by \eqref{eq:parameterizationofg}. The $m$ index in \cite{matrixelements} corresponds to the quantum number $m$ in \eqref{eq:whittakerwavefunction}. The conventions of \cite{matrixelements} imply that the $\rho$ index must be fixed to $\sqrt{2}$ so that the matrix element satisfies \eqref{eq:restriction}.\footnote{That is, the eigenvalue of $G_2 - G_1$ is given by $\frac{\rho^2}{2}\cdot(\text{sign } \rho)$.} Given $\mu$, we must restrict $m \in \mathbb{Z} + \mu$ so that the central element has the correct eigenvalue. Note that \cite{matrixelements} only considers $\mu = 0$ or $\frac{1}{2}$, which are the only values relevant for $SL(2,\mathbb{R})$ representations. The normalization of \eqref{eq:whittakerwavefunction} can be checked explicitly \cite{whittakerfunctionorthogonality}.

Since $D^{s,\mu}_{m \rho}(g)$ corresponds to the state \eqref{eq:whittakerwavefunction}, the projector \eqref{eq:projector} is given, up to normalization, by
\begin{equation}
\sum_{m \in \mathbb{Z} + \mu} D^{s,\mu}_{m \rho}(g_2) \left(D^{s,\mu}_{m \rho}(g_1)\right)^* = \sum_{m \in \mathbb{Z} + \mu} D^{s,\mu}_{\rho m }(g_1^{-1}) D^{s,\mu}_{m \rho}(g_2) = D^{s,\mu}_{\rho \rho}(g_1^{-1}g_2),
\end{equation}
where we have used the fact that $D^{s,\mu}_{m \rho}$ is a matrix element of a unitary \sltr representation. Again, we must plug in $\rho = \sqrt{2}$. Thus, to determine \eqref{eq:projector}, we need to know matrix elements of the principal series representations of \sltr in the basis that diagonalizes $G_2 - G_1$. These matrix elements are given for the $\mu = 0$ and $\mu = \frac{1}{2}$ cases in (4.21) of \cite{matrixelements}. The matrix elements for other values of $\mu$ are a linear combination of the matrix elements for $\mu = 0$ and $\mu = \frac{1}{2}$. This is because the Casimir operator \eqref{eq:casimir} is a second-order differential operator, so general solutions of the eigenvalue equation for the Casimir may be written as a linear combination of two linearly independent solutions. The coefficients of the linear combination are chosen by requiring that
\begin{align}
\begin{split}
D^{s,\mu}_{\rho \, \rho^\prime}(1) &= \delta\left((\text{sign } \rho) \cdot\frac{\rho^2}{2} - (\text{sign } \rho^\prime) \cdot\frac{(\rho^\prime)^2}{2}\right), \\
D^{s,\mu}_{\rho \, \rho^\prime}(-1) &= \delta\left((\text{sign } \rho) \cdot\frac{\rho^2}{2} - (\text{sign } \rho^\prime) \cdot\frac{(\rho^\prime)^2}{2}\right) e^{2 \pi i \mu},
\end{split}
\end{align}
where $1$ denotes the identity element of \sltr and $-1$ denotes the central element. The normalization of $D^{s,\mu}_{\rho \rho}(g_1^{-1} g_2)$ differs from the normalization of \eqref{eq:projector} by the well-known Plancherel measure, which is explained further in \cite{SL2RNotes} and \cite{finestructure}, Appendix C.

Finally, we explain how we obtain \eqref{eq:rhof} starting from \eqref{eq:imageoffactorizationmap}. First, it is obvious that \eqref{eq:rhof} must vanish for $|T_1 - T_2| > 2 \pi$, given that $\rho_f$ is the reduced density matrix of a wavefunction of the form \eqref{eq:imageoffactorizationmap}. So, we only need to determine $\rho_f$ for $|T_1 - T_2| < 2\pi$. As a trick, we compactify $T \sim T + 4 \pi$, which does not change the result for $|T_1 - T_2| < 2\pi$. With this compactification, only the $\mu = 0$ and $\mu = \frac{1}{2}$ representations are relevant. We may rewrite \eqref{eq:imageoffactorizationmap} as
\begin{equation}
\label{eq:psiscompactify}
\begin{split}
&\Psi_s(T_L,\chi_L,T_R,\chi_R) \\
&=\frac{\sqrt{2 s \sinh 2 \pi s}}{2 \pi^2} \frac{e^{i(e^{\chi_R} - e^{\chi_L})\tan\left(\frac{T_L - T_R}{2}\right)}}{\left|\cos\left(\frac{T_L - T_R}{2}\right)\right|}  K_{2 i s}\left(2 e^{\frac{\chi_L + \chi_R}{2}} \left|\sec \left(\frac{T_L - T_R}{2}\right)\right|\right)\frac{(1 + \text{sign } (\cos(\frac{T_R - T_L}{2})))}{2}.
\end{split}
\end{equation}
Using (4.21) of \cite{matrixelements}, we can obtain \eqref{eq:psiscompactify} as a linear combination of the matrix elements $D^{s,0}_{\rho \rho^\prime}(g_L^{-1}e^{\pi G_1}g_R)$ and $D^{s,\frac{1}{2}}_{\rho \rho^\prime}(g_L^{-1}e^{\pi G_1}g_R)$ with $\rho = \sqrt{2}$ and $\rho^\prime = - \sqrt{2}$.\footnote{In \cite{matrixelements}, $\mu$ is denoted by $\epsilon$ and $k = \frac{1}{2} + i s$.} To get an expression involving Whittaker functions, we may decompose
\begin{equation}
D^{s,\mu}_{\rho \rho^\prime}(g_L^{-1}e^{\pi G_1}g_R) = \sum_{m \in \mathbb{Z} + \mu} D^{s,\mu}_{\rho m}(g_L^{-1})D^{s,\mu}_{m \rho^\prime}(e^{\pi G_1}g_R).
\end{equation}
We then find that \eqref{eq:psiscompactify} is given by
\begin{equation}
\label{eq:psiscompactifymsbasis}
\begin{split}
&\Psi_s(T_L,\chi_L,T_R,\chi_R)\\
&= \sum_{m \in \mathbb{Z}} \frac{1}{2\sqrt{s \tanh \pi s}} \braket{T_L \, \chi_L |s \,\, -m} \braket{T_R \, \chi_R|s \, m} + \sum_{m \in \mathbb{Z} + \frac{1}{2}} \frac{1}{2\sqrt{s \coth \pi s}} \braket{T_L \, \chi_L | s \,\, -m} \braket{T_R \, \chi_R | s \, m},
\end{split}
\end{equation}
with $\braket{T \, \chi | s \, m}$ still defined as in \eqref{eq:whittakerwavefunction}. Compactifying $T$ causes the inner product to be modified:
\begin{equation}
\braket{s^\prime \, m^\prime | s \, m} = 2\delta(s^\prime - s)\delta_{m m^\prime}.
\end{equation}
The (unnormalized) reduced density matrix of $\ket{f} \equiv \int_0^\infty ds f(s) \ket{\Psi_s}$ with $\ket{\Psi_s}$ given in \eqref{eq:psiscompactifymsbasis} is then
\begin{equation}
\label{eq:reduceddensitymatrixofpsiscompactified}
\text{Tr}_L \ket{f}\bra{f}
= \int_0^\infty ds \, |f(s)|^2\left(\sum_{m \in \mathbb{Z}} \frac{1}{2 s \tanh \pi s} \ket{s \, m} \bra{s \, m} + \sum_{m \in \mathbb{Z} + \frac{1}{2}} \frac{1}{2 s \coth \pi s} \ket{s \, m} \bra{s \, m}\right).
\end{equation}
The matrix element \eqref{eq:rhofmtrixelement} follows from using \eqref{eq:projector} to convert \eqref{eq:reduceddensitymatrixofpsiscompactified} to the $T,\chi$ basis. Using \eqref{eq:projector} again, we obtain \eqref{eq:rhof} from \eqref{eq:rhofmtrixelement}.

\section{The Euclidean Schwarzian path integral}
\label{sec:euclidpathintegral}

In this section, we provide more details on the Euclidean Schwarzian path integral that appears in \cite{stanfordwitten,zhenbinyang}. We also describe an alternate way to compute \eqref{eq:rhohh} by analytically continuing the path integral with the insertion of a ``full defect.'' While in this case there are subtleties in analytically continuing the Euclidean path integral, our result is consistent with \eqref{eq:rhohh}.

The path integral (or propagator) of the Euclidean Schwarzian theory is given in first-order form as follows:
\begin{equation}
\label{eq:Euclideanschwarzianpathintegral}
\begin{split}
&K_\beta(\phi_2,\psi_2,\phi_1,\psi_1) \equiv
\\
&\int \mathcal{D}\phi \mathcal{D}\psi \mathcal{D}\pi_\phi \mathcal{D}\pi_\psi  \exp\left(\int_0^\beta d\tau \left[ i \pi_\psi \psi^\prime +  i \pi_\phi \phi^\prime - \frac{1}{2 \phi_b} \left[ \frac{\pi_\psi^2}{2} + i \pi_\phi e^\psi - \frac{1}{2} e^{2 \psi}\right]   \right]\right),
\end{split}
\end{equation}
where at time $\tau = 0$ the fields $\phi,\psi$ are fixed to $\phi_1,\psi_1$ and at time $\tau = \beta$ the fields are fixed to $\phi_2,\psi_2$. Upon integrating out all of the fields except $\phi$, we obtain a Euclidean path integral with action
\begin{equation} I = 2 \phi_b \int_0^\beta d\tau \, \left(  \frac{1}{2} \left(\frac{\phi^{\prime \prime}}{\phi^\prime}\right)^2  - \frac{1}{2} ( \phi^\prime)^2  \right),  \end{equation}
and a reparameterization-invariant measure for $\phi$. The canonical form of \eqref{eq:Euclideanschwarzianpathintegral} indicates that the propagator satisfies Schrodinger's equation:
\begin{equation}
\label{eq:Euclideanshrodingereq}
\frac{1}{2 \phi_b} \left[ -\frac{\partial_\psi^2}{2} +   e^\psi \partial_\phi - \frac{1}{2} e^{2 \psi}\right]K_\beta(\phi,\psi,\phi_1,\psi_1) = -\partial_\beta K_\beta(\phi,\psi,\phi_1,\psi_1).
\end{equation}

A perturbative analysis of \eqref{eq:Euclideanschwarzianpathintegral} about the classical solution $\phi(\tau) = \phi_1 + (\phi_2 - \phi_1)\frac{\tau}{\beta}$ has instabilities at one-loop when $\phi_2 > \phi_1 + 2 \pi$. Thus, we do not attempt to make sense of $\eqref{eq:Euclideanschwarzianpathintegral}$ for $\phi_2 > \phi_1 + 2 \pi$.\footnote{A more thorough analysis for the case $\phi_2 > \phi_1 + 2\pi$ would be interesting to study in future work. See \cite{JTdefects} for further comments on winding numbers greater than one.} For $\phi_2 < \phi_1$, \eqref{eq:Euclideanschwarzianpathintegral} vanishes because no off-shell configurations satisfy the constraint $\phi^\prime = \frac{e^\psi}{2 \phi_b}$, which is enforced with a Lagrange multiplier.

The propagator \eqref{eq:Euclideanschwarzianpathintegral} has already been solved in \cite{zhenbinyang}, equation (5.35). The expression there is given in Poincar\'e variables $x,z$, where $x$ is Euclidean Poincar\'e time and $z$ is a rescaled spatial coordinate. The Poincar\'e variables are related to $\phi,\psi$ as follows:
\begin{equation} x = \tan \frac{\phi}{2}, \quad z = \frac{e^{\psi}}{2}  \sec^2 \frac{\phi}{2}. \end{equation}
Applying this change of variables to (5.35) of \cite{zhenbinyang}, we obtain
\begin{equation}
\label{eq:propfromzhenbin}
\begin{split}
\tilde{K}_\beta(\phi_2,\psi_2,\phi_1,\psi_1) = \exp\left((e^{\psi_2} + e^{\psi_1})\cot\left(\frac{\phi_1 - \phi_2}{2}\right) - e^{\psi_2} \tan \frac{\phi_2}{2} + e^{\psi_1} \tan \frac{\phi_1}{2}\right) \\
\times  \frac{1}{\pi^2} \frac{1}{|\sin \frac{\phi_1 - \phi_2}{2}|} \int_{0}^{\infty}ds s \sinh(2 \pi s) e^{-\frac{s^2}{2}\beta} K_{2is}\left(\frac{ 2 e^{\frac{\psi_2 + \psi_1}{2}}}{|\sin \frac{\phi_1 - \phi_2}{2}|}\right),
\end{split}
\end{equation}
and the measure associated with this propagator is $d\phi \, d\psi$. In \cite{zhenbinyang}, the propagator was used to compute exact correlators of operators inserted on the boundary of the disk. For $\phi_1,\phi_3 \in (-\pi,\pi)$ satisfying $\phi_3 > \phi_1$, the propagator \eqref{eq:propfromzhenbin} satisfies the gluing property:
\begin{equation}
\label{eq:gluing}
\int_{\phi_1}^{\phi_3} d\phi_2 \int_{-\infty}^\infty d\psi_2 \, \tilde{K}_{\beta_1}(\phi_3,\psi_3,\phi_2,\psi_2)\tilde{K}_{\beta_2}(\phi_2,\psi_2,\phi_1,\psi_1) = \tilde{K}_{(\beta_1 + \beta_2)}(\phi_3,\psi_3,\phi_1,\psi_1).
\end{equation}
The correctness of \eqref{eq:gluing} follows from the self-consistency of equation (6.74) of \cite{zhenbinyang} upon setting one of the operators to the identity.

To obtain \eqref{eq:Euclideanschwarzianpathintegral}, we must remove the factor of $e^{ -e^{\psi_2} \tan \frac{\phi_2}{2} + e^{\psi_1} \tan \frac{\phi_1}{2}}$ from \eqref{eq:propfromzhenbin}, which does not affect the gluing property in \eqref{eq:gluing}. Thus, we conclude that for $2 \pi > \phi_2 - \phi_1 > 0$, \eqref{eq:Euclideanschwarzianpathintegral} is given by
\begin{equation}
\label{eq:Euclideanschwarzianpropagator}
\begin{split}
K_\beta(\phi_2,\psi_2,\phi_1,\psi_1) = \exp\left((e^{\psi_2} + e^{\psi_1})\cot\left(\frac{\phi_1 - \phi_2}{2}\right) \right) \\
\times  \frac{1}{\pi^2} \frac{1}{\sin \frac{\phi_2 - \phi_1}{2}} \int_{0}^{\infty}ds s \sinh(2 \pi s) e^{-\frac{s^2}{4 \phi_b}\beta} K_{2is}\left(\frac{ 2 e^{\frac{\psi_2 + \psi_1}{2}}}{\sin \frac{\phi_2 - \phi_1}{2}}\right),
\end{split}
\end{equation}
where we have restored the variable $\phi_b$. Note that \eqref{eq:Euclideanschwarzianpropagator} satisfies Schrodinger's equation \eqref{eq:Euclideanshrodingereq} and only depends on $\phi_1$ and $\phi_2$ through their difference, $\phi_2 - \phi_1$.

While the Hartle-Hawking density matrix \eqref{eq:rhohh} follows from \eqref{eq:HHwavefunction}, we can also directly relate \eqref{eq:rhohh} to the analytic continuation of the Euclidean path integral. We consider \eqref{eq:Euclideanschwarzianpropagator} with the insertion of a ``full defect:''
\begin{equation} K_\beta(\phi_2 + 2 \pi,\psi_2,\phi_1,\psi_1) = \frac{1}{\pi^2} \int_0^\infty ds \, e^{- \beta \frac{s^2}{4 \phi_b}} \, s \sinh(2 \pi s)  \frac{e^{(e^{\psi_1} + e^{\psi_2})\cot \frac{\phi_1 - \phi_2}{2}}}{\sin \frac{\phi_1 - \phi_2}{2}} K_{2 i s}\left(\frac{2 e^{\frac{\psi_1 + \psi_2}{2}}}{\sin \frac{\phi_1 - \phi_2}{2}}\right). \label{eq:propwithfulldefect} \end{equation}
As explained earlier, \eqref{eq:propwithfulldefect} is only valid for $0 < \phi_1  - \phi_2 < 2 \pi$. To analytically continue, we rotate $\phi_1 \rightarrow i \phi_x$ and $\phi_2 \rightarrow i \phi_y$ while keeping $0 < \phi_x - \phi_y < 2 \pi$. We also multiply \eqref{eq:propwithfulldefect} by a factor of $i$ because the measure changes under analytic continuation: $d \phi \, d\psi \rightarrow i \, d\phi \, d\psi$. The result is
\begin{equation}
\label{eq:propagatorwithfulldefectanalyticallycontinued}
i K_\beta(i\phi_y + 2 \pi,\psi_y,i\phi_x,\psi_x) = \int_0^\infty ds \, e^{- \beta \frac{s^2}{4 \phi_b}} \, s \sinh(2 \pi s)  \frac{e^{- \pi s}}{2 \pi}\frac{i e^{i(e^{\psi_x} + e^{\psi_y})\coth \frac{\phi_y - \phi_x}{2}}}{\sinh \frac{\phi_x - \phi_y}{2}} H^{(1)}_{2 i s}\left(\frac{2 e^{\frac{\psi_x + \psi_y}{2}}}{\sinh \frac{\phi_x - \phi_y}{2}}\right). \end{equation}
While we originally assumed that $0 < \phi_x - \phi_y < 2 \pi$, we see that \eqref{eq:propagatorwithfulldefectanalyticallycontinued} is nonsingular in the extended domain of $\phi_x > \phi_y$. We may analytically continue \eqref{eq:propagatorwithfulldefectanalyticallycontinued} to this extended domain. We may interpret \eqref{eq:propagatorwithfulldefectanalyticallycontinued} as a matrix element of the unnormalized density matrix $\tilde{\rho}_\beta$ that corresponds to the Hartle-Hawking state. For $\phi_x > \phi_y$, we have
\begin{equation}
\label{eq:matrixelementagreaterthanb}
\begin{split}
&\braket{\phi_y \, \psi_y|\tilde{\rho}_\beta|\phi_x \, \psi_x} = \\
&\int_0^\infty ds \, e^{- \beta \frac{s^2}{4 \phi_b}} \, s \sinh(2 \pi s)  \frac{e^{- \pi s}}{2 \pi}\frac{i e^{i(e^{\psi_x} + e^{\psi_y})\coth \frac{\phi_y - \phi_x}{2}}}{\sinh \frac{\phi_x - \phi_y}{2}} H^{(1)}_{2 i s}\left(\frac{2 e^{\frac{\psi_x + \psi_y}{2}}}{\sinh \frac{\phi_x - \phi_y}{2}}\right) \times e^{-i e^{\psi_y} \tanh \phi_y + i e^{\psi_x}\tanh \phi_x }.
\end{split}
\end{equation}
To be consistent with the extra phases introduced in \eqref{eq:propwithhalfdefectanalyticallycontinued}, we have introduced additional phases in \eqref{eq:matrixelementagreaterthanb}. Using \eqref{eq:rindlertoglobaltime} to convert to global time variables, we obtain
\begin{equation}
\label{eq:matrixelementagreaterthanbglobaltime}
\begin{split}
\braket{T_y \, \chi_y | \tilde{\rho}_\beta | T_x \, \chi_x} &= \int_0^\infty ds \,  s \sinh (2 \pi s) e^{-\beta \frac{s^2}{4 \phi_b}} \left(\frac{i e^{i(e^{\chi_x} + e^{\chi_y})\cot\left(\frac{T_y - T_x}{2}\right)}}{2 \pi}  \csc\left(\frac{T_x - T_y}{2}\right)\right) \\
 &\times
e^{- \pi s} H^{(1)}_{2 i s}\left(2 e^{\frac{\chi_x + \chi_y}{2}} \csc\left(\frac{T_x - T_y}{2}\right)\right), \quad 2\pi > T_x - T_y > 0.
\end{split}
\end{equation}
While we have derived \eqref{eq:matrixelementagreaterthanbglobaltime} from an expression in Rindler variables, we have analytically continued \eqref{eq:matrixelementagreaterthanbglobaltime} to the full domain $2 \pi > T_x - T_y > 0$ where it continues to be nonsingular.

To obtain an expression for $\braket{T_y \, \chi_y | \tilde{\rho}_\beta | T_x \, \chi_x}$ in the domain $2 \pi > T_y - T_x > 0$, one can start with \eqref{eq:matrixelementagreaterthanbglobaltime} and impose the condition that $\tilde{\rho}_\beta$ is Hermitian. One cannot analytically continue from the region $2 \pi > T_x - T_y > 0$ to the region $2 \pi > T_y - T_x > 0$. If we assume that $\tilde{\rho}_\beta$ is the reduced density matrix of a wavefunction of the form \eqref{eq:mostgeneralpsi}, then we may conclude that \eqref{eq:matrixelementagreaterthanbglobaltime} vanishes for $|T_x - T_y| > 2 \pi$, which, after normalizing $\tilde{\rho}_\beta$, results in agreement with \eqref{eq:rhofmtrixelement} for $|f(s)|^2 = \frac{1}{N_\beta} 2 s \sinh(2 \pi s) e^{- \beta \frac{s^2}{4 \phi_b}}$.


\begin{thebibliography}{99}

\bibitem{RT}
S. Ryu and T. Takayanagi, ``Holographic Derivation of Entanglement Entropy from AdS/CFT,'' Phys. Rev. Lett. {\bf 96}, 181602 (2006) [arXiv:hep-th/0603001].
	
\bibitem{HRT}
V. E. Hubeny, M. Rangamani, and T. Takayanagi, ``A Covariant Holographic Entanglement Entropy Proposal,'' JHEP {\bf 0707}, 62 (2007) [arXiv:0705.0016].
	
\bibitem{LM}
A. Lewkowycz and J. Maldacena, ``Generalized gravitational entropy,'' JHEP {\bf 1308}, 90  (2013) [arXiv:1304.4926].
	
\bibitem{FLM}
T. Faulkner, A. Lewkowycz, and J. Maldacena, ``Quantum corrections to holographic entanglement entropy,'' JHEP {\bf 1311}, 74 (2013) [arXiv:1307.2892].

\bibitem{EW}
N. Engelhardt and A. C. Wall, ``Decoding the Apparent Horizon: A Coarse-Grained Holographic Entropy,'' [arXiv:1706.02038].

\bibitem{alliteration}
X. Dong and A. Lewkowycz, ``Entropy, Extremality, Euclidean Variations, and the Equations of Motion,'' Phys. Rev. Lett. {\bf 121}, 211301  (2018) [arXiv:1705.08453].

\bibitem{QuantumPenrose}
R. Bousso, A. Shahbazi-Moghaddam, and M. Tomasevic, ``Quantum Information Bound on the Energy'', [arXiv:1909.02001].

\bibitem{SurfaceTheory}
N.~Engelhardt and S.~Fischetti, ``Surface Theory: the Classical, the Quantum, and the Holographic,'' arXiv:1904.08423.

\bibitem{pagecurve1}
A. Almheiri, N. Engelhardt, D. Marolf, and H. Maxfield, ``The entropy of bulk quantum fields and the entanglement wedge of an evaporating black hole,'' [arXiv:1905.08762].

\bibitem{pagecurve2}
G. Penington, ``Entanglement Wedge Reconstruction and the Information Paradox,'' [arXiv:1905.08255].

\bibitem{quantumfocussing}
R.~Bousso, Z.~Fisher, S.~Leichenauer, and A. C. Wall, ``A Quantum Focussing Conjecture,'' Phys. Rev. D {\bf 93}, 064044 (2016) [arXiv:1506.02669].

\bibitem{JLMS}
D. L. Jafferis, A. Lewkowycz, J. Maldacena, S. J. Suh, ``Relative entropy equals bulk relative entropy,'' JHEP {\bf 1606}, 4 (2016) [arXiv:1512.06431].

\bibitem{HarlowRT}
D. Harlow, ``The Ryu-Takayanagi Formula from Quantum Error Correction,'' Comm. Math. Phys. {\bf 354}, 865 (2017) [arXiv:1607.03901].

\bibitem{alphabits}
P. Hayden and G. Penington, ``Learning the Alpha-bits of Black Holes,'' [arXiv:1807.06041].

\bibitem{universalrecovery}
J. Cotler, P. Hayden, G. Penington, G. Salton, B. Swingle, M. Walter, ``Entanglement Wedge Reconstruction via Universal Recovery Channels'', Phys. Rev. X {\bf 9}, 031011 (2019) [arXiv:1704.05839].

\bibitem{freegravitons}
V. Benedetti and H. Casini, ``Entanglement entropy of linearized gravitons in a sphere,'' [arXiv:1908.01800].

\bibitem{livingontheedge}
W. Donnelly, B. Michel, D. Marolf and J. Wien, ``Living on the Edge: A Toy Model for Holographic Reconstruction of Algebras with Centers,'' JHEP {\bf 1704}, 93 (2017) [arXiv:1611.05841].

\bibitem{Jackiw}
R. Jackiw, ``Lower Dimensional Gravity,'' Nucl. Phys. {\bf B252}, 343 (1985).

\bibitem{Teitelboim}
C. Teitelboim, ``Gravitation and Hamiltonian Structure in Two Space-Time Dimensions,'' Phys. Lett. {\bf 126B}, 343 (1983).

\bibitem{MSY}
J. Maldacena, D. Stanford, and Z. Yang, ``Conformal symmetry and its breaking in two dimensional Nearly Anti-de-Sitter space,'' [arXiv:1606.01857].

\bibitem{FactorizationProblem}
D.~Harlow and D.~Jafferis, ``The Factorization Problem in Jackiw-Teitelboim Gravity,'' [arXiv:1804.01081].

\bibitem{Tachikawa}
K. Ohmori and Y. Tachikawa, ``Physics at the entangling surface,'' J. Stat. Mech. {\bf 1504}, P04010 (2015) [arXiv:1406.4167].

\bibitem{DJ}
T. Dumitrescu and D. Jafferis, work in progress.

\bibitem{commentsonSYK}
J. Maldacena and D. Stanford, ``Comments on the Sachdev-Ye-Kitaev model,'' Phys. Rev. D {\bf 94}, 106002 (2016)  [arXiv:1604.07818].

\bibitem{eternaltraversable}
J. Maldacena and X.-L. Qi, ``Eternal traversable wormhole,'' [arXiv:1804.00491].

\bibitem{symmetriesnearhorizon}
H. W. Lin, J. Maldacena and Y. Zhao, ``Symmetries Near the Horizon,'' JHEP {\bf 1908}, 49 (2019) [arXiv:1904.12820].

\bibitem{stanfordwitten}
D.~Stanford and E.~Witten, ``Fermionic Localization of the Schwarzian Theory,'' JHEP {\bf 1710}, 8 (2017) [arXiv:1703.04612].

\bibitem{clocksandrods}
A. Blommaert, T. G. Mertens, and H. Verschelde, ``Clocks and Rods in Jackiw-Teitelboim Quantum Gravity,'' JHEP {\bf 1909}, 60 (2019) [arXiv:1902.11194].

\bibitem{holographiccomplexity}
K. Goto, H. Marrochio, R. C. Myers, L. Queimada, B. Yoshida, ``Holographic Complexity Equals Which Action?'' JHEP {\bf 1902}, 160 (2019) [arXiv:1901.00014].

\bibitem{mertensbhevaporation}
T. G. Mertens, ``Towards Black Hole Evaporation in Jackiw-Teitelboim Gravity,'' JHEP {\bf 1907}, 97 (2019) [arXiv:1903.10485].

\bibitem{sss}
P. Saad, S. H. Shenker and D. Stanford, ``JT gravity as a matrix integral,'' [arXiv:1903.11115].

\bibitem{semiclassicalramp}
P. Saad, S. H. Shenker and D. Stanford, ``A semiclassical ramp in SYK and in gravity,'' [arXiv:1806.06840].

\bibitem{saad}
P. Saad, ``Late Time Correlation Functions, Baby Universes, and ETH in JT Gravity,'' [arXiv:1910.10311].













\bibitem{JenniferLin2}
J. Lin, ``Ryu-Takayanagi Area as an Entanglement Edge Term,'' [arXiv:1704.07763].

\bibitem{JenniferLin}
J. Lin, ``Entanglement entropy in Jackiw-Teitelboim Gravity,'' [arXiv:1807.06575].

\bibitem{JTdefects}
T. G. Mertens and G. J. Turiaci, ``Defects in Jackiw-Teitelboim Quantum Gravity,'' JHEP {\bf 1908}, 127 (2019) [arXiv:1904.05228].

\bibitem{Wong}
W. Donnelly and G. Wong, ``Entanglement branes, modular flow, and extended topological quantum field theory,'' [arXiv:1811.10785].

\bibitem{casinihuertarosabal}
H. Casini, M. Huerta, and J. A. Rosabal, ``Remarks on entanglement entropy for gauge fields,'' 	Phys. Rev. D {\bf 89}, 085012 (2014) [arXiv:1312.1183].

\bibitem{commentsonEE}
J. Lin and D. Radicevic, ``Comments on Defining Entanglement Entropy,'' [arXiv:1808.05939].

\bibitem{zhenbinyang}
Z.~Yang, ``The Quantum Gravity Dynamics of Near Extremal Black Holes,'' JHEP {\bf 1905}, 205 (2019) [arXiv:1809.08647].

\bibitem{GabrielWong}
G. Wong, ``A note on entanglement edge modes in Chern Simons theory,'' JHEP {\bf 1808}, 020 (2018) [arXiv:1706.04666].

\bibitem{Rangamanietal}
X. Dong, A. Lewkowycz, and M. Rangamani, ``Deriving covariant holographic entanglement,'' JHEP {\bf 1611}, 028 (2016) [arXiv:1607.07506].


\bibitem{SL2RNotes}
A.~Kitaev, ``Notes on \sltr representations,'' [arXiv:1711.08169].


\bibitem{MorsePotential}
P.~Zhang, ``Morse Potential, Contour Integrals, and Asian Options,'' [arXiv:1010.3820].



\bibitem{yangmills2d}
W.~Donnelly, ``Entanglement entropy and nonabelian gauge symmetry'', Class. Quantum Grav. {\bf 31} no. 21, 214003 (2014) [arXiv:1406.7304].

\bibitem{exactquantization}
L. V. Iliesiu, S. S. Pufu, H. Verlinde and Y. Wang, ``An exact quantization of Jackiw-Teitelboim gravity,'' [arXiv:1905.02726].

\bibitem{kitaevsuh}
A. Kitaev, S. J. Suh, ``Statistical mechanics of a two-dimensional black hole,'' JHEP {\bf 1905}, 198 (2019) [arXiv:1808.07032].

\bibitem{Szmytkowski}
R. Szmytkowski, ``On the derivative of the Legendre function of the first kind with respect to its degree,'' J. Phys. A: Math. Gen. {\bf 39} (2006) 15147.
	
\bibitem{matrixelements}
D.~Basu and K.~B.~Wolf, ``The unitary irreducible representations of $SL(2,\mathbb{R})$ in all subgroup reductions,'' J. Math. Phys. {\bf 23} (1982) 189.

\bibitem{whittakerfunctionorthogonality}
R. Szmytkowski and S. Bielski, ``An orthogonality relation for the Whittaker functions of the second kind of imaginary order,'' [arXiv:0910.1492].

\bibitem{finestructure}
A. Blommaert, T. G. Mertens and H. Verschelde, ``Fine Structure of Jackiw-Teitelboim Quantum Gravity,'' JHEP {\bf 1909}, 66 (2019) [arXiv:1812.00918].

\bibitem{DonnellyFridel}
W. Donnelly and L. Freidel, ``Local subsystems in gauge theory and gravity,'' JHEP {\bf 1609}, 102 (2016) [arXiv:1601.04744].



\bibitem{TTbarQM}
D. J. Gross, J. Kruthoff, A. Rolph and E. Shaghoulian, ``$T\bar{T}$  in AdS$_2$ and Quantum Mechanics,'' [arXiv:1907.04873].

\end{thebibliography}
\end{document}